\documentclass[a4paper,preprint,12pt]{elsarticle}
\usepackage[english]{babel} 
\usepackage[latin1]{inputenc}  
\usepackage{amssymb}  
\usepackage{amsmath}    
\usepackage{pb-diagram}
\usepackage{epstopdf}        
\usepackage{graphicx}      
 \usepackage{epsfig}    
\usepackage{color}
\usepackage{fancyhdr}  
 \usepackage{natbib}  
\usepackage{rotating}
\usepackage{pdflscape}

\usepackage[T1]{fontenc}        % accents dans le DVI
\usepackage{url}                % citer des adresses �lectroniques et des URL
\usepackage{mathtools}                 %      write over a line
\usepackage{vmargin}    
\usepackage[title]{appendix}
\usepackage{lineno}
\usepackage{setspace}
\usepackage{color,soul}
\setulcolor{red} % set underlining color
\setstcolor{blue}   % set overstriking color
\sethlcolor{red}  % set hig
\biboptions{square} 
  
\newcommand{\bu}{\bullet}     
\newcommand{\bo}[1]{\mathbf{#1}} 
\def\indic{\hbox{1\kern-.24em\hbox{I}}}      % indicatrice

\newcommand{\var}{\mathbb{V}}  
\newcommand{\esp}{\mathbb{E}}    
\newcommand{\trace}{\text{Tr}}         
  
\newcommand{\khi}{\mathcal{X}}   
\newcommand{\x}{x}
\newcommand{\X}{X}

\newcommand{\Z}{Z} 
\newcommand{\R}{\mathbb{R}}
\newcommand{\proba}{\mathbb{P}} 

\newcommand{\M}{f}

\newcommand{\T}{T}    
\newcommand{\N}{\mathbb{N}}      
\newcommand{\NN}{n}

\newcommand{\norme}[1]{\left|\left| #1 \right|\right|_{L^2}}

\newcommand{\normf}[1]{\left|\left| #1 \right|\right|_F}

\newcommand{\D}{\Sigma}

\newtheorem{prop}{Proposition}{\bf}{\it} 
\newtheorem{defi}{Definition}{\bf}{\it}  

\newtheorem{theorem}{Theorem}{\bf}{\it}     
\newtheorem{lemma}{Lemma}{\bf}{\it}          
\newtheorem{rem}{Remark}{\bf}{\it} 
\newtheorem{corollary}{Corollary}{\bf}{\it} 
{\bf}{\it}

\sethlcolor{green}      

\catcode`\"=\active 
\catcode`\"=\active 
\def"{\og\ignorespaces}
\def"{{\fg}} 
\usepackage{etoolbox}
\makeatletter
\patchcmd{\ps@pprintTitle}{\footnotesize\itshape
       Preprint submitted to \ifx\@journal\@empty Elsevier
       \else\@journal\fi\hfill\today}{\relax}{}{}
\makeatother
         
\begin{document}
  
\begin{frontmatter}  
  
%-----------  
\title{Efficient dependency models for some distributions}             
\author[a,b]{Matieyendou Lamboni\footnote{Corresponding author: matieyendou.lamboni[at]gmail.com; Submitted to MCS on March 01, 2021}}     
\address[a]{University of Guyane, Department DFRST, 97346 Cayenne, French Guiana, France}
\address[b]{228-UMR Espace-Dev, University of Guyane, University of R\'eunion, IRD, University of Montpellier, France.}     
                   
%%-----------------------------------------------------                  
%------------------------------------------------------ 
\begin{abstract}
Dependency functions of dependent variables are relevant for i) performing uncertainty quantification and sensitivity analysis in presence of dependent variables and/or correlated variables, and ii) simulating random dependent variables. In this paper, we mathematically derive practical dependency functions for classical multivariate distributions such as Dirichlet, elliptical distributions and independent uniform (resp. gamma and Gaussian) variables under constraints that are ready to be used. Since such dependency models are used for sampling random values and we have many dependency models for every joint cumulative distribution function, we provide a way for choosing the efficient sampling function using multivariate sensitivity analysis. We illustrate our approach by means of numerical simulations.
\end{abstract}                    
                        
\begin{keyword}                 
Dependent generalized sensitivity indices \sep Dependent variables \sep  Efficient sampling \sep Multivariate distributions \sep Random values.
\end{keyword}    
               
\end{frontmatter}      
% \setpagewiselinenumbers     
%\modulolinenumbers[1]   
%\linenumbers
%\doublespacing                                
                   
% REQUIRED                
% \begin{AMS}      
%  65C20, 62H20, 49Q12, 65C50, 65C60.                           
% \end{AMS}  

%  49Q12       SA              62P30   Appli in enge
% 62F12 Asymptotic properties of estimators
% 78M05 method of moments
% 60E15 Inequalities; stochastic orderings   
% 39B72 Systems of functional equations and inequalities    
% 26D10 Inequalities involvving derivatives and differential and integral  
%65C50 Other computational problems in probability computational: 
%65C60 Computational problems in statistics
%62H20 Measures of association (correlation, canonical correlation, etc.)
%65C20 Models, numerical methods [See also]68U20
%65D15 Algorithms for functional approximation

%%%%%%%%%%%%%%%%%%%%%%%%%%%%%%%%%%%%%%%%%%%%%    
\section{Introduction}
%-----------------------context  
Multivariate conditional quantile transform such as the inverse Rosenblatt transformations (\cite{obrien75,skorohod76,arjas78,ruschendorf81}) and the conditional distribution method (\cite{devroye86,mcneil15}) imply regression representations of random vectors (\cite{skorohod76,ruschendorf93,ruschendorf09}), which also imply dependency models of random vectors (\cite{lamboni21}). A dependency model (DM) expresses a subset of dependent variables as a function of the remaining variable(s) and new independent variables. For instance, a random vector of three dependent variables can be written as follows: 
$$
(\X_1,\, \X_2,\, \X_3) \stackrel{d}{=} r_1\left(\X_1, \, Z_2,\, Z_3 \right) \, , 
$$ 
where $r_1$ is a dependency function, and $\X_1, \, Z_2,\, Z_3$ are independent variables. \\ 

Using the above example, we see that such dependency function eases the computation of the partial derivatives of $\X_2$ and $\X_3$ w.r.t. $\X_1$.
Moreover, dependency functions and the inverse dependency functions are relevant for performing uncertainty quantification and sensitivity analysis in presence of dependent variables (including correlated variables) known as dependent multivariate sensitivity analysis (dMSA), dependent derivative global sensitivity measures (dDGSM) and dependent elementary effects (dEE) (\cite{lamboni21}). Indeed, such functions are used for i) simulating random dependent variables and then combining with models for performing dMSA, dDGSM and dEE, and ii) computing partial derivatives of functions with non-independent variables (\cite{lamboni21ecm}). \\ 
       
Some generic DMs have been provided in \cite{skorohod76} for two variables and in \cite{lamboni21,lamboni21ar} for every dimension by making use of the cumulative distribution functions of dependent variables. Most of such distribution-based DMs require some mathematical derivations before being able to be used.    
In this paper, we derive practical and generic DMs that are ready to be used for some classical and well-known distributions such as elliptical distributions for correlated variables, generalized Dirichlet distributions,  
 uniformly distributed variables under constraints, independent gamma variables under constraints, independent Gaussian variables under constraints and related distributions. As the DM is not unique for a given $d$-dimensional random vector, we propose a methodology for choosing the efficient DM by making use of multivariate sensitivity analysis (MSA) (\cite{lamboni18a, lamboni11, lamboni09,gamboa14,lamboni19,lamboni22}), including Sobol' indices (\cite{sobol93,saltelli00}).\\
  
From now on, the paper is organized as follows: in Section \ref{sec:pdm}, we give  the practical DMs for some well-known distributions, and Section \ref{sec:stra} provides a sensitivity-based approach for identifying the efficient DM when supplementary information such as the causality is not available. Section \ref{sec:num} deals with the implementation of our approach by means of numerical simulations. We conclude this work in Section \ref{sec:con}. 
						    				         	       	          	                        
%%%%%%%%%%%%%%%%%%%%%%%%%%%%%%%%%%%%%%%%%%%%%%%%          
\section*{General notation}              
%---                              
For integer $d >0$, let $\bo{\X} :=(X_1, \, \ldots,\, X_d)$ be a vector of input variables. For $j \in \{1,\,\ldots,\, d\}$, we use $\bo{\X}_{\sim j}:=(\X_i,\, i \in \{1,\,\ldots,\, d\}\backslash \{j\})$. Thus, we have the partition $\bo{\X} =(\X_{j},\, \bo{\X}_{\sim j})$. 
We use $\bo{Z} \stackrel{d}{=}(\X_{j},\, \bo{\X}_{\sim j})$ to say that $\bo{Z}$ and $(\X_{j},\, \bo{\X}_{\sim j})$ have the same distribution. For $\bo{a} \in\R^\NN$, we use $\norme{\bo{a}}$ for the Euclidean norm. For a matrix $\Sigma \in \R^{\NN\times \NN}$, we use $\trace(\Sigma)$ for the trace of $\Sigma$, and $\normf{\Sigma} :=\sqrt{\trace\left(\Sigma \Sigma^\T \right)}$ for the Frobenius norm of $\Sigma$. We use $\esp[\cdot]$ for the expectation and $\var[\cdot]$ for the variance.
 
%%%%%%%%%%%%%%%%%%%%%%%%%%%%%%%%%%%%%%%%%%%%    
%%-------------------------------------------------------------
%%--------------------------------------------------
\section{Dependency models for some multivariate distributions} \label{sec:pdm}
%-------           
In this section, we provide generic and practical expressions of dependency models for i) elliptical distributions such as Student distributions, ii) Dirichlet distribution and independent uniformly distributed variables under constraints, iii) independent gamma variables under constraints, iv) independent Gaussian distributions under constraints and related distributions.\\

Namely, the vector of $d$ dependent variables $\bo{\X}$ has $F$ as the joint cumulative distribution function (CDF) and  $F_{x_j}$ or $F_j$ as marginal CDF with $j=1,\ldots,\, d$. We use $F_{j}^{\leftarrow}$ for the generalized inverse of $F_j$ and $F_{k | j}$ for the conditional CDF of $\X_k$ given $\X_j$. Let $\bo{Z} \sim \mathcal{U}(0,\,1)^{d-1}$ be $d-1$ independent variables following the standard uniform distribution, and $(w_1, \ldots, w_{d-1}) := (\sim j)$ be any permutation of the set $\{1,\ldots, d\}\setminus \{j\}$. A generic DM of $(\X_j, \, \bo{\X}_{\sim j})$ is given by (\cite{skorohod76,lamboni21}).  

\begin{equation} \label{eq:depmpr}
\bo{\X}_{\sim j} \stackrel{d}{=} r_j\left(\X_j, \, \bo{Z} \right) \, , 
\end{equation} 
where $r_j : \R^{d} \to \R^{d-1}$ is a measurable dependency function, and $\X_j$ is independent of  $\bo{Z}$. A generic expression of $r_j =: (r_{w_{1}}, \ldots, r_{w_{d-1}})$ is given by (\cite{skorohod76,lamboni21})
%-- 
\begin{eqnarray}     \label{eq:genmodelr}                           
r_{w_{1}}(\X_j, \, Z_1)   & = & F_{w_1| j}^{\leftarrow} \left(\Z_1 \,|\, \X_j \right) \nonumber \\ r_{w_{2}}(\X_j, \, Z_1, \, Z_2)  & = &  F_{w_2 | j, w_1}^{\leftarrow}\left(\Z_2 \,|\, \X_j,\, r_{w_{1}}(\X_j, \, Z_1) \right)   \\     
 & \vdots &  \nonumber \\          
r_{w_{d-1}}(\X_j, \, \bo{Z}) &=&  F_{w_{d-1} | j, \bo{w}_{\sim d-1}}^{\leftarrow} \left(\Z_{d-1} \,|\, \X_j, r_{w_{1}}(\X_j, \, Z_1), \ldots, r_{w_{d-2}}(\X_j, \, \bo{Z}_{\sim d-1}) \right) \, .          \nonumber 
\end{eqnarray}    
   
From Equation (\ref{eq:genmodelr}), it is clear that for any distribution, we need to derive the analytical expressions of the conditional CDFs to obtain $r_j$. It is worth noting that the dependency function is not unique, and sometime we obtain simple expressions of $r_j$ using other distributions of $\bo{Z}$ rather than the uniform distribution (\cite{lamboni21}). Thus, we are going to provide the expressions of $r_j$ for some distributions and for any $d \in \N^*$ by making use of any distribution of $\bo{Z}$ (consisted of independent variables).
  
%%---   
\begin{rem} \label{rem:lcdepm}
Let $\bo{\X} =(\X_j, \bo{\X}_{\sim j})$ with $\bo{\X}_{\sim j} =r_j\left(\X_j, \bo{Z} \right)$, $\bo{Y} \stackrel{d}{=} \Sigma \bo{\X} + \boldsymbol{\mu}$ with $\Sigma \in \R^{d \times d}$ a left-triangular matrix and $\boldsymbol{\mu} \in \R^d$. Then, a DM of $\bo{Y}$ is 
\begin{equation} \label{eq:lcdepm}
\left(Y_j,\, \bo{Y}_{\sim j}\right) = \Sigma \, \left[\frac{Y_j-\mu_j}{\Sigma_{jj}}, r_j\left(\frac{Y_j-\mu_j}{\Sigma_{jj}},\, \bo{Z} \right)^\T \right]^\T + \boldsymbol{\mu} \, . 
\end{equation} 
\end{rem}     
             
%%-----------------------------------  
\subsection{Elliptical distributions: correlated variables}
%------ 
 Elliptical distributions (\cite{muirhead82,fang90,embrechts02}) are often used for modeling correlated variables. This section provides DMs for some of such distributions.\\

Formally, for a symmetric and positive definite matrix $\Sigma \in \R^{d \times d}$, we use $\mathcal{L}$ for the Cholesky factor of $\Sigma$, that is, the lower triangular matrix of the Cholesky decomposition of $\Sigma=\mathcal{L}\mathcal{L}^\T$. For a vector $\bo{v} \in \R^d$, we use $diag(\bo{v}) \in \R^{d \times d}$ for a diagonal matrix. We also use $\mathcal{N}_d(\boldsymbol{\mu}, \, \Sigma)$ for the multivariate Gaussian distribution.

%-------------------------                              
\begin{prop} \label{prop:gauco} 
 Let $\Phi$ be the CDF of the standard Gaussian distribution,  $\mathcal{L}$ be the Cholesky factor of $\Sigma$, $\Sigma_{\sim j} = diag(\Sigma_{w_i w_i},\, i=1,\ldots, d-1)$ and $\bo{Z} \sim \mathcal{N}_{d-1}\left(\boldsymbol{\mu}_{\sim j}, \, \Sigma_{\sim j} \right)$.\\  
             
If  $(\X_{j},\, \bo{\X}_{\sim j}) \sim \mathcal{N}_d \left(\boldsymbol{\mu}, \Sigma \right)$, then a DM of  $\bo{\X}$ is given by 
\begin{eqnarray} \label{eqn:gdis}        
\displaystyle  
 \left(\X_j, \bo{\X}_{\sim j}\right) & \stackrel{}{=} &  \mathcal{L} \; 
         \left[\begin{array}{c} \frac{1}{\sqrt{\Sigma_{j j}}}\left(\X_{j} -\esp[\X_{j}]\right) \\
	 \Sigma_{\sim j}^{-1/2}	\left(\bo{Z} - \boldsymbol{\mu}_{\sim j} \right)\\
							\end{array}        
			\right]
			+   \boldsymbol{\mu}   \, .                           
\end{eqnarray}              
\end{prop}          
\begin{preuve}
It is an extension of the DM provided in \cite{lamboni21}. 
\begin{flushright}  
$\Box$    
\end{flushright} 
\end{preuve}

 The above DM of $\bo{\X}$ can be used for generating the random  values of a multivariate $t$-distribution (\cite{kotz04}), as a $t$-distribution (i.e., $\bo{t}_d(\nu, \boldsymbol{\mu}, \Sigma)$) is a mixture of a multivariate Gaussian distribution (e.g., $\bo{R} \sim \mathcal{N}_d(\boldsymbol{0}, \, \Sigma)$) and an inverse gamma distribution (i.e., $W \sim Ig(\nu/2, \nu/2)$). Indeed, when $\bo{R}$ is independent of $ W$, we have
$$   
\bo{Y} \stackrel{d}{:=} \sqrt{W} \bo{R} + \boldsymbol{\mu}  \sim \bo{t}_d(\nu, \boldsymbol{\mu}, \Sigma) \, .
$$      
However, the above representation of a $t$-distribution is far away from a DM for a multivariate $t$-distribution. Therefore, we give a DM for a $t$-distribution in Proposition \ref{prop:stco}. We use $t(\nu,\, 0,\, 1)$ for the standard $t$-distribution having $\nu >2$ degrees of freedom and $T_\nu$ for its CDF.    
  
%-------------------------                                    
\begin{prop} \label{prop:stco}   
Let $\bo{Z} :=\left(Z_{w_i} \sim t(\nu + i,\, 0,\, 1),\, i=1,\,\ldots,\, d-1 \right)$ be a vector of  independent variables, $\mathcal{L}$ be the Cholesky factor of $\Sigma$ and $\bo{Z}$ be independent of $\X_j$. \\ 
 
 If  $(\X_{j},\, \bo{\X}_{\sim j}) \sim \bo{t}_d(\nu, \boldsymbol{\mu}, \Sigma)$, then a DM of $\bo{X}$ is given by 
\begin{equation} \label{eq:stcor}    
\displaystyle    
(\X_j, \bo{\X}_{\sim j})   
	\stackrel{d}{=}  \mathcal{L} 
					\left[	\begin{array}{c}   
							\frac{\X_{j} -\mu_{j}}{\sqrt{\Sigma_{j j}}} \\
 \sqrt{\frac{\nu \Sigma_{j j} + \left(\X_{j} -\mu_{j}\right)^2}{\Sigma_{j j}(\nu +1)}} Z_{w_1}  \\
  \vdots \\             
 \sqrt{\frac{\left(\nu \Sigma_{j j} + \left(\X_{j} -\mu_{j}\right)^2 \right) \prod_{k=1}^{d-2}\left(\nu + k+ Z_{w_k}^2 \right)}{\Sigma_{j j} \prod_{k=1}^{d-1}(\nu +k)}} Z_{w_{d-1}} \\
							\end{array}  \right] + \boldsymbol{\mu} \, .          
\end{equation}                          
\end{prop}  
\begin{preuve}  
See Appendix \ref{app:theo:stco}.  
\begin{flushright}     
$\Box$ 
\end{flushright}     
\end{preuve}      

When the degree of freedom $\nu=1$, the standard $t$-distribution  $t(\nu=1,\, 0,\, 1)$ is also known as the standard Cauchy distribution, that is, $C(0,\, 1) \stackrel{d}{=} t(\nu=1,\, 0,\, 1)$. We use $C_d(\boldsymbol{\mu},\, \Sigma)$ for a multivariate Cauchy distribution, and we provide a DM for such distribution in Proposition \ref{prop:cauco}.
          
%-------------------------                                  
\begin{prop} \label{prop:cauco}  
Let $T_1$ be the CDF of the standard Cauchy distribution; $Z_{w_i} \sim t(1+i,\, 0,\, 1)$ with $i=1,\ldots, d-1$ and $\X_j$ be independent variables. \\  

 If  $(\X_{j},\, \bo{\X}_{\sim j}) \sim C_d(\boldsymbol{0}, \mathcal{I})$, then a DM of $\bo{\X}$ is given by 
%--------   
\begin{equation} \label{eq:cauunc} 
\displaystyle  
\begin{array}{cccl} 
 \X_{w_1} &=&  h_1\frac{Z_{w_1}}{\sqrt{2}}\, ,  & \quad h_1:= \sqrt{1 + \X_{j}} \\
 \X_{w_2} &=& h_2 \frac{Z_{w_2}}{\sqrt{3}} \, ,  & \quad h_2 :=\sqrt{h_1\left(h_1+ \frac{Z_{w_1}}{\sqrt{2}}\right)}   \\ 
  &  & \vdots  & \\  
\X_{w_{d-1}} &=&   h_{w_{d-1}} \frac{Z_{w_{d-1}}}{\sqrt{d}}\, , & \quad  h_{w_{d-1}} :=\sqrt{h_{w_{d-2}}\left(h_{w_{d-2}} + \frac{Z_{w_{d-2}}}{\sqrt{d-1}}\right)}   \\ 
\end{array} \, .      
\end{equation}
\end{prop}  
\begin{preuve} 
See Appendix \ref{app:theo:cauco}. 
\begin{flushright}         
$\Box$
\end{flushright}  
\end{preuve}
      
Using the dependency function provided by Equation (\ref{eq:cauunc}), we are able to derive a DM of $\bo{Y} \sim C_d(\boldsymbol{\mu},\, \Sigma)$. We just have to apply the identity given by Equation (\ref{eq:lcdepm}) (see Remark \ref{rem:lcdepm}) because we known that $\bo{Y} \sim C_d(\boldsymbol{\mu},\, \Sigma) \Longleftrightarrow \bo{Y} \stackrel{d}{=} \mathcal{L}\,  C_d(\boldsymbol{0},\, \mathcal{I}) + \boldsymbol{\mu}$ with $\mathcal{L}$ the Cholesky factor of $\Sigma$.

%%%%%%%%%%%%%%%%%%%%%%%%%%%%%%%%%%%%%%%%%%%%%%%%%%%%%%%%%%%%%%% 
%%-----------------------------------    
\subsection{Dependency model for the Dirichlet distribution and its generalization}
%------    
Consider the parameters $\bo{a}, \bo{b} \in \R^d_+$ and the $d$-dimensional regular polytope and simplex given by, respectively,  
$$
S_d =\left\{\bo{\x} \in (0,\,1)^d : \,  \sum_{k=1}^d \x_k < 1 \right\},
\qquad 
S_d^* =\left\{\bo{\x} \in (0,\,1)^d : \,  \sum_{k=1}^d \x_k = 1 \right\}  \, .
$$ 

The generalized Dirichlet distribution, that is, $GD\left(\bo{a},\, \bo{b}\right)$ has the density (\cite{connor69})
$$
\rho\left(\bo{\x}\right) = \frac{1}{D} \prod_{k=1}^d \x_k^{a_k-1} \left(1- \sum_{i=1}^k \x_i \right)^{b_{k}-1}\times \indic_{S_d}(\bo{\x}) \, ,  
$$    
where $\indic_{S_d}(\bo{\x}) =1$ if $\bo{\x} \in S_d$ and $0$ otherwise;
and $D$ is a constant ensuring that $\int_{\R^d} \rho\left(\bo{\x}\right)\, d\bo{\x} =1$. It is known that (\cite{connor69,chang10}) 
\begin{eqnarray} \label{eq:bindg}  
& & \X_j \sim Beta\left(a_j, b_j + \sum_{\substack{k=1,\, k\neq j}}^{d}(a_k+b_k-1) \right) \, , \nonumber \\
& & Z_j :=\frac{\X_j}{1- \sum_{k=1}^{j-1}\X_k} \sim Beta\left(a_j, \, b_j + \sum_{\substack{k=j+1}}^{d} (a_k +b_k -1)\right) \, ,     \nonumber 
\end{eqnarray}  
where $Z_j$ with $j=1,\ldots, d$ are mutually independent and  beta-distributed. It is to be noted that we obtain the Dirichlet distribution when $\bo{b} =(1,\ldots, 1,\, b_d =: a_{d+1})$ with $a_{d+1}>0$.  
Putting all these elements together, Lemma \ref{lem:diri} provides a dependency model for the generalized Dirichlet distribution. For $j \in\{1, \ldots, d\}$, recall that $(w_1,\ldots, w_{d-1})$ is any permutation of $\{1, \ldots, d\}\setminus\{j\}$. 
%---   
%-------------------------        
\begin{lemma} \label{lem:diri} (Chang et al. \cite{chang10})  \\                
%--
Let $Z_{w_i} \sim Beta\left(a_{w_i}, \,  b_{w_i} + \sum_{\substack{k=i+1}}^{d-1} (a_{w_k} + b_{w_k} -1) \right)$ with $i=1,\ldots, d-1$ and $\X_{j} \sim Beta\left(a_{j}, \, b_j + \sum_{\substack{k=1}}^{d-1} (a_{w_k} + b_{w_k} -1)\right)$ be independent variables. \\  
 
If $(\X_{j},\,\bo{\X}_{\sim j}) \sim GD(a_{j},\, a_{w_1}, \ldots, a_{w_{d-1}}, \, b_{j},\, b_{w_1}, \ldots, b_{w_{d-1}})$, then    
%--- 
\begin{equation} \label{eq:dirig}
\bo{\X}_{\sim j} \stackrel{d}{=} \left(Z_{w_1} \left(1-\X_{j}\right), \ldots, Z_{w_{d-1}} \left(1-\X_{j}\right) \prod_{k=1}^{d-2}\left(1-Z_{w_k} \right) \right)  \, . 
\end{equation}               
\end{lemma}

\begin{rem}  
In order to work with the initial marginal CDFs, we can replace in (\ref{eq:dirig}) $Z_{w_i}$ with $F_{Z_{w_i}}^{-1}\left(F_{\X_{w_i}'}(\X_{w_i}') \right)$ where $\X_{w_i}' \sim Beta\left(a_{w_i}, \, b_{w_i} + \sum_{\substack{k=1,\, k\neq w_i}}^{d} (a_{k} + b_{k} -1)\right)$, $F_{\X_{w_i}'}$  is the CDF of $\X_{w_i}'$and  $F_{Z_{w_i}}$ is the CDF of $Z_{w_i}$. 
\end{rem} 

For the Dirichlet distribution with parameters $(a_1, \ldots, a_{d+1}) \in\R^{d+1}_+$, that is, $D\left(a_1,\, \ldots, \, a_{d+1}\right)$, Equation (\ref{eq:dirig}) comes down to the following model. If $(\X_{j},\,\bo{\X}_{\sim j}) \sim D(a_{j},\, a_{w_1}, \ldots, a_{w_{d-1}}, \, a_{d+1})$, then we have    
%--- 
\begin{equation} \label{eq:diri}
\bo{\X}_{\sim j} \stackrel{d}{=} \left(Z_{w_1} \left(1-\X_{j}\right), \ldots, Z_{w_{d-1}} \left(1-\X_{j}\right) \prod_{k=1}^{d-2}\left(1-Z_{w_k} \right) \right)  \, , 
\end{equation} 
where $\X_{j} \sim Beta\left(a_{j}, \, \sum_{\substack{k=1}}^{d-1} a_{w_k} + a_{d+1}\right)$ and $Z_{w_i} \sim Beta\left(a_{w_i}, \, \sum_{\substack{k=i+1}}^{d-1} a_{w_k} + a_{d+1} \right)$ with $i=1,\ldots, d-1$  are mutually independent.  \\ 

Further extension of the Dirichlet distribution known as the $p$-Generalized Dirichlet distribution has been proposed in \cite{barthe10}. Formally, let us consider the $d$-dimensional unit $p$-ball and $p$-simplex given by 
%--   
\begin{equation} \label{eq:lpball}
B_{d, p} := \left\{\bo{x} \in \R^d :  \sum_{k=1}^d |x_k|^p <1  \right\},   
\qquad 
B_{d, p}^* := \left\{\bo{x} \in \R^d :   \sum_{k=1}^d |x_k|^p = 1  \right\} 
 \, ,     
\end{equation}  
which generalize $S_d$ and $S_d^*$, respectively. The random  vector $\bo{\X}$ follows the $p$-Generalized Dirichlet distribution on $B_{d, p}$ (i.e., $\bo{\X} \sim p\!-\!DG\left(\bo{a},\, \bo{b} \right)$) whenever (\cite{barthe10}) 
%----  
$$ 
\bo{\X} \stackrel{d}{=} \left(R_1 Y_1^{1/p}, \ldots, R_d Y_d^{1/p} \right) \, , 
$$  
where $(Y_1, \ldots, Y_d) \sim GD\left(\bo{a},\, \bo{b} \right)$ is independent of $(R_1, \ldots, R_d)$, and $R_i$ with $i=1, \ldots, d$ are independent and Rademacher distributed, that is, the probability $\proba(R_i=1) =\proba(R_i=-1) =1/2$. We provide a dependency function of $\bo{\X}$ in Proposition \ref{prop:dirigp}. 

%-------------------------       
\begin{prop} \label{prop:dirigp}              
%--
Let $Z_{w_i} \sim Beta\left(a_{w_i}, \,  b_{w_i} +\sum_{\substack{k=i+1}}^{d-1} (a_{w_k} + b_{w_k}-1) \right)$ with $i=1, \ldots, d-1$; $R_i$ with $i=1,\ldots, d$ and $Z_{j} \sim Beta\left(a_{j}, \, b_j + \sum_{\substack{k=1}}^{d-1} (a_{w_k} + b_{w_k} -1)\right)$ be independent variables, and  $\X_{j} \stackrel{d}{=} R_j Z_j^{1/p}$. \\  
 
If $(\X_{j},\,\bo{\X}_{\sim j}) \sim p\!-\!DG\left(a_{j},\, a_{w_1}, \ldots, a_{w_{d-1}}, \, b_{j},\, b_{w_1}, \ldots, b_{w_{d-1}} \right)$, then
%--- 
\begin{equation} \label{eq:dirigp}
\bo{\X}_{\sim j} \stackrel{d}{=} \left(R_{w_1} \left[Z_{w_1} \left(1- |\X_{j}|^p\right)\right]^{1/p}, \ldots, R_{w_{d-1}} \left[Z_{w_{d-1}}\left(1-|\X_{j}|^p\right) \prod_{k=1}^{d-2}\left(1-Z_{w_k} \right)\right]^{1/p} \right)  \, .     
\end{equation}  
\end{prop} 
\begin{preuve}
See Appendix \ref{app:prop:dirigp}. 
\begin{flushright}
$ \Box$  
\end{flushright} 
\end{preuve}

\begin{rem}
Similar derivation has been provided in \cite{barthe10} with different parameters for the beta-distributed variables.
\end{rem}

Using (\ref{eq:dirigp}), we can deduce a dependency function of $(|\X_j|, \, |\X_{w_{1}}|,  \ldots, |\X_{w_{d-1}}|)$ with $\bo{\X} \sim p\!-\!DG\left(\bo{a},\, \bo{b} \right)$ because $(|\X_j|, \, |\X_{w_{1}}|,  \ldots, |\X_{w_{d-1}}|) \stackrel{d}{=} \left( Y_1^{1/p}, \ldots, Y_d^{1/p} \right)$ (see Appendix \ref{app:prop:dirigp}). Moreover, we are able to provide a DM for the $p$-Generalized Dirichlet distribution on $B_{d, p}^*$ given by (\ref{eq:lpball}) based on the following definition. 
%-----  
\begin{defi}
A random vector $\bo{\X}$ has the $p$-Generalized Dirichlet distribution on  $B_{d, p}^*$ if for all $u \subset \{1, \ldots, d\}$ with $|u|=d-1$, 
$$
\bo{\X}_u \stackrel{d}{=} \left(R_1 Y_1^{1/p}, \ldots, R_{d-1} Y_{d-1}^{1/p} \right) \, , 
$$     
where $(Y_1, \ldots, Y_{d-1}) \sim GD\left(\bo{a},\, \bo{b} \right)$ on $B_{d-1, p}$.
\end{defi} 
 
The derivation of a DM of $\bo{\X}$ following the $p$-Generalized Dirichlet distribution on  $B_{d, p}^*$ is straightforward, and it is given in Proposition \ref{prop:dirigpeq}. To that end, we use $\bo{w} := (w_1,\ldots, w_{d-1})$ for an arbitrary permutation of $\{1, \ldots, d\}\setminus \{j\}$.    
 
%-------------------------        
\begin{prop} \label{prop:dirigpeq}       
%--
Let $Z_{w_i} \sim Beta\left(a_{w_i}, \,  b_{w_i} +\sum_{\substack{k=i+1}}^{d-2} (a_{w_k} + b_{w_k}-1) \right)$ with $i=1, \ldots, d-2$; $R_i$ with $i=1,\ldots, d$ and $Z_{j} \sim Beta\left(a_{j}, \, b_j + \sum_{\substack{k=1}}^{d-2} (a_{w_k} + b_{w_k} -1)\right)$ be independent variables, and  $\X_{j} \stackrel{d}{=} R_j Z_j^{1/p}$. \\  
 
If $\left(\X_{j},\,\bo{\X}_{\bo{w}\setminus w_{d-1}},\, \X_{w_{d-1}} \right)  \sim p\!-\!DG\left(a_{j},\, a_{w_1}, \ldots, a_{w_{d-1}}, \, b_{j},\, b_{w_1}, \ldots, b_{w_{d-1}} \right)$ on $B_{d, p}^*$, then
%--- 
\begin{eqnarray} \label{eq:dirigpeq} 
\bo{\X}_{\bo{w}\setminus w_{d-1}} &\stackrel{d}{=}& \left(R_{w_1} \left[Z_{w_1}(1- |\X_{j}|^p)\right]^{1/p}, \ldots, R_{w_{d-2}} \left[Z_{w_{d-2}} \left(1-|\X_{j}|^p\right) \prod_{k=1}^{d-3}\left(1-Z_{w_k} \right)\right]^{1/p} \right) \, ,  \nonumber \\   
\X_{w_{d-1}} &\stackrel{d}{=}& R_{w_{d-1}}\left[1 -\sum_{i \in \left\{j, \bo{w}\setminus \{w_{d-1}\} \right\}} |\X_i|^p  \right]^{1/p} \, .      
\end{eqnarray}  
\end{prop}

%%%%%%%%%%%%%%%%%%%%%%%%%%%%%%%%%%%%%%%%%%%%%%%%%%%%%%%%%%%%%%%  
%%-----------------------------------     
\subsection{Independent uniform random variables under constraints}
%------   
In this section, we provide the DMs for independent uniform variables under some equality and inequality constraints and related distributions. \\ 

A random vector $\bo{\X} = \left(X_1, \ldots, \X_d \right)$ is said uniformly distrusted over the simplex $S_d^*$ (i.e., $\bo{\X} \sim \mathcal{U}(S_d^*)$) if for any Borel-measurable set $B$, the probability $\proba\left(\bo{\X} \in B \right) = \lambda(B)/\lambda(S_d^*)$ with $\lambda$ the Lebesgue measure (\cite{marsaglia61,devroye86}). It is known that $\left(\X_{1}, \ldots, \X_{d}, \X_{d+1} \right) \sim \mathcal{U}(S_d^*)$ implies that $\left(\X_{1}, \ldots, \X_{d} \right) \sim \mathcal{U}(S_d)$ (see \cite{marsaglia61}). The uniform distribution on the unit $p$-ball, that is, $\mathcal{U}(B_{d, p})$ with $B_{d, p}$ given by (\ref{eq:lpball}) extends the uniform distribution over $S_d$ (i.e., $\mathcal{U}(S_d)$) (see \cite{barthe10}). If $\bo{\X} \sim \mathcal{U}(B_{d, p})$, then we have (see \cite{barthe10})
$$ 
\bo{\X} \sim p\!-\!D\left(p^{-1}, \ldots, p^{-1},\, 1 \right) \, , 
$$ 
where $p\!-\!D\left(p^{-1}, \ldots, p^{-1},\, 1 \right)$ stands for $p\!-\!GD\left(\bo{a},\, \bo{b} \right)$ with  $\bo{a} =\left(p^{-1}, \ldots, p^{-1}\right)$ and $\bo{b} = (1, \ldots, 1)$. We then deduce the following DMs using Proposition~\ref{prop:dirigp}.

%-------             
\begin{prop} \label{prop:dirigp2}  
Let $Z_{w_i} \sim Beta\left(p^{-1}, \, p^{-1}(d-i-1) +1 \right)$ with $i=1, \ldots, d-1$; $R_i$ with $i=1,\ldots, d$ and $Z_{j} \sim Beta\left(p^{-1}, \, (d-1)p^{-1} + 1 \right)$ be independent variables. \\ 
   
If $(\X_{j},\,\bo{\X}_{\sim j}) \sim \mathcal{U}(B_{d, p})$, then $\X_{j} \stackrel{d}{=} R_j Z_j^{1/p}$ and          
%--- 
\begin{equation} \label{eq:dirigp2} 
\bo{\X}_{\sim j} \stackrel{d}{=} \left(R_{w_1} \left[Z_{w_1}(1- |\X_{j}|^p)\right]^{1/p}, \ldots, R_{w_{d-1}} \left[Z_{w_{d-1}}\left(1-|\X_{j}|^p\right) \prod_{k=1}^{d-2}\left(1-Z_{w_k} \right)\right]^{1/p} \right)  \, .       
\end{equation}
\end{prop}   
 
The derivation of a DM for the uniform distribution over the regular polytope (i.e., $\bo{\X} \sim \mathcal{U}(S_d)$)) is a particular case of Proposition \ref{prop:dirigp2}, as $x_i \in (-1,\, 1)$ implies $|x_i| \in (0, \, 1)$. Using the DM of $p\!-\!D\left(p^{-1}, \ldots, p^{-1},\, 1 \right)$ given by (\ref{eq:dirigp2}) and taking the absolute value of that model, we obtain a DM of the uniformly distributed random vector on $B_{d, p}^+ :=\left\{\bo{x} \in \R^d_+ : \sum_{i=1}^d x_i^p < 1 \right\}$, that is, $\bo{\X} \sim \mathcal{U}(B_{d, p}^+)$ (see Proposition \ref{prop:unigp}).  \\ 
%------- 
\begin{prop} \label{prop:unigp}
Let $Z_{w_i} \sim Beta\left(p^{-1}, \, p^{-1}(d-i-1) +1 \right)$ with $i=1, \ldots, d-1$ and $Z_{j} \sim Beta\left(p^{-1}, \, (d-1)p^{-1} + 1 \right)$ be independent variables, and  $\X_{j} \stackrel{d}{=} Z_j^{1/p}$. \\ 

If $\left(\X_{j}, \, \bo{\X}_{\sim j} \right) \sim \mathcal{U}(B_{d, p}^+)$, then
\begin{equation} \label{eq:uni1}
\displaystyle                                                        
\bo{\X}_{\sim j} \stackrel{d}{=} \left(\left[Z_{w_1}(1- |\X_{j}|^p)\right]^{1/p}, \ldots, \left[Z_{w_{d-1}} \left(1-|\X_{j}|^p\right) \prod_{k=1}^{d-2}\left(1-Z_{w_k} \right)\right]^{1/p} \right)\, .
\end{equation} 
\end{prop}         
 
When $p=1$, we obtain a DM of $\bo{\X} \sim \mathcal{U}(S_d)$. In the same sense, the DMs of $\bo{\X} \sim \mathcal{U}(S_d^*)$, $\bo{\X} \sim \mathcal{U}(B_{d, p}^*)$ and $\bo{\X} \sim \mathcal{U}(B_{d, p}^{*+})$ can be obtained using Proposition \ref{prop:dirigpeq}. We give such results in Proposition \ref{prop:univeq}.

\begin{prop} \label{prop:univeq}
Let $Z_{w_i} \sim Beta\left(p^{-1}, \, p^{-1}(d-i-2) +1 \right)$ with $i=1, \ldots, d-2$; $R_i$ with $i=1,\ldots, d$ and $Z_{j} \sim Beta\left(p^{-1}, \, (d-2)p^{-1} + 1 \right)$ be independent variables. \\
 
$\quad$ (i) If $\left(\X_{j},\,\bo{\X}_{\bo{w} \setminus w_{d-1}},\, \X_{w_{d-1}} \right) \sim  \mathcal{U}(B_{d, p}^*)$, then $\X_{j} \stackrel{d}{=} R_j Z_j^{1/p}$ and 
%--- 
\begin{eqnarray} \label{eq:univ2} 
\bo{\X}_{\bo{w} \setminus w_{d-1}} &\stackrel{d}{=}& \left(R_{w_1} \left[Z_{w_1}(1- |\X_{j}|^p)\right]^{1/p}, \ldots, R_{w_{d-2}} \left[Z_{w_{d-2}} \left(1-|\X_{j}|^p\right) \prod_{k=1}^{d-3}\left(1-Z_{w_k} \right)\right]^{1/p} \right) \, ,  \nonumber \\
\X_{w_{d-1}} &\stackrel{d}{=}& R_{w_{d-1}}\left[1 -\sum_{i \in \left\{j, \bo{w} \setminus \{w_{d-1}\} \right\}} |\X_i|^p  \right]^{1/p} \, .  
\end{eqnarray}        

$\quad$ (ii) If $\left(\X_{j},\,\bo{\X}_{\bo{w} \setminus w_{d-1}},\, \X_{w_{d-1}} \right)  \sim  \mathcal{U}(B_{d, p}^{*+})$, then $\X_{j} \stackrel{d}{=} Z_j^{1/p}$ and 
%--- 
\begin{eqnarray} \label{eq:univ3} 
\bo{\X}_{\bo{w} \setminus w_{d-1}} &\stackrel{d}{=}& \left(\left[Z_{w_1}(1- |\X_{j}|^p)\right]^{1/p}, \ldots, \left[ Z_{w_{d-2}} \left(1-|\X_{j}|^p\right) \prod_{k=1}^{d-3}\left(1-Z_{w_k} \right)\right]^{1/p} \right) \, ,  \nonumber \\
\X_{w_{d-1}} &\stackrel{d}{=}& \left[1 -\sum_{i \in \left\{j, \bo{w} \setminus \{w_{d-1}\} \right\}} \X_i^p  \right]^{1/p} \, .        
\end{eqnarray}  
\end{prop}
   
When $p=1$, Equation (\ref{eq:univ3}) provides a DM of the uniformly distributed random vector over the simplex  $S_{d}^*$. 

%%-----------------------------------      
\subsection{Independent gamma random variables under constraints}
%------     
This section provides DMs of independent variables $\bo{\X}$ following the gamma distribution under constraints such as $\sum_{k=1}^d \X_k = c$ and $\sum_{k=1}^d \X_k < c$ with $c>0$. We also provide DMs for a wide class of distributions related to the former distributions. Corollaries \ref{coro:idconstga}-\ref{coro:idconstga2} give such results.\\

Namely, for $\bo{a} \in \R^{d}_+$ and $\beta>0$, we use $\Gamma(a_j, \, \beta)$ for a gamma distribution having  $G_j$ as CDF with $j =1, \ldots, d$ and $B1\left(c,\, a, \, b \right)$ for the beta distribution of first-kind (\cite{mcdonald95}). We also use $\left\{\X_j \sim \Gamma(a_j, \, \beta),\, j=1,\ldots, d\, :\, \sum_{j=1}^d \X_j = c \right\}$ for  independent gamma variables  subject to $\sum_{j=1}^d \X_j = c$ and $F_i,\, i=1, \ldots, d$ for continuous CDFs. 
                
%--------------------------------------              
\begin{corollary} \label{coro:idconstga}     
Let $Z_{w_i} \sim Beta\left(a_{w_i}, \, \sum_{\substack{k=i+1}}^{d-1} a_{w_k}\right)$ and $Z_{j} \sim B1\left(c,\, a_{j}, \, \sum_{\substack{k=1}}^{d-1} a_{w_k} \right)$ with $i=1,\ldots, d-2$ be independent variables. \\ 

$\quad$ (i) If  $\left(\X_{j}^c, \bo{\X}_{\sim j}^c \right) \stackrel{d}{=} \left\{\X_i \sim \Gamma(a_i, \, \beta),\, i=j, \, w_1, \ldots, w_{d-1}\, :\, \sum_{i=1}^d \X_i = c \right\}$, then we have  $\X_{j}^c \stackrel{d}{=}  Z_{j}$, 
\begin{equation} \label{eq:ga} 
\displaystyle                                                         
\begin{array}{rcl}            
\X_{w_1}^c   & = &  Z_{w_1} \left(c-\X_{j}^c \right) \\     
  &\vdots &  \\          
\X_{w_{d-2}}^c  &=&  Z_{w_{d-2}} \left(c-\X_{j}^c\right) \prod_{k=1}^{d-3}\left(1-Z_{w_k} \right) \\
 \X_{w_{d-1}}^c &=&  \left(c-\X_{j}^c\right) \prod_{k=1}^{d-2}\left(1-Z_{w_k} \right) \\
	\end{array} \, .           
\end{equation}
  
$\quad$ (ii) If $\left(\X_{j}^c, \bo{\X}_{\sim j}^c \right) \stackrel{d}{=} \left\{\X_i \sim F_i,\, i=j, \, w_1, \ldots, w_{d-1}\, :\, \sum_{i=1}^d  G_i^{-1}\left(F_{i}(\X_i) \right)= c \right\}$, then we have    $\X_{j}^c \stackrel{d}{=} F_{j}^{-1}\left(G_{j} ( Z_{j})\right)$,     
\begin{equation} \label{eq:gacop} 
\displaystyle 
\begin{array}{rcl}             
\X_{w_1}^c   & = & F_{w_1}^{-1}\left(G_{w_1} \left( Z_{w_1} \left(c-  G_{j}^{-1} \left(F_{j}\left(\X_{j}^c \right)\right) \right) \right) \right) \\
  &\vdots &  \\          
\X_{w_{d-2}}^c  &=&  F_{w_{d-2}}^{-1}\left(G_{w_{d-2}} \left( Z_{w_{d-2}} \left(c-  G_{j}^{-1} \left(F_{j}\left(\X_{j}^c \right)\right)\right) \prod_{k=1}^{d-3}\left(1-Z_{w_k} \right) \right) \right)  \\
 \X_{w_{d-1}}^c &=&  F_{w_{d-1}}^{-1}\left(G_{w_{d-1}} \left(  \left(c- G_{j}^{-1} \left(F_{j}\left(\X_{j}^c \right)\right)\right) \prod_{k=1}^{d-2}\left(1-Z_{w_k} \right) \right) \right)  \\
	\end{array} \,  .  
\end{equation}     
\end{corollary} 
\begin{preuve} 
See Appendix \ref{app:theo:idconstga}.   
\begin{flushright} 
$\Box$ 
\end{flushright}
\end{preuve}     

While Corollary \ref{coro:idconstga} provides  DMs for the gamma distribution under some equality constraints, Corollary \ref{coro:idconstga2} gives the results for the same distribution under inequality constraints.
   
%-------------------------                                     
\begin{corollary} \label{coro:idconstga2}   
Let $Z_{w_i} \sim Beta\left(a_{w_i}, \, \sum_{\substack{k=i+1}}^{d-1} a_{w_k} +1 \right),\, i=1,\,\ldots,\, d-1$ and  $Z_{j} \sim B1\left(c,\,a_{j}, \, \sum_{\substack{k=1}}^{d-1} a_{w_k} +1\right)$ be independent variables, $F_i$ be a continuous CDF. \\ 

 $\quad$ (i) If $\left(\X_{j}^c, \bo{\X}_{\sim j}^c \right) \stackrel{d}{=} \left\{\X_i \sim \Gamma(a_i, \, \beta),\, i=j, \, w_1, \ldots, w_{d-1}\, :\, \sum_{i=1}^d \X_i < c \right\}$, then we have $\X_{j}^c \stackrel{d}{=} \, Z_{j}$,
\begin{equation} \label{eq:ga2} 
\displaystyle                                                        
\begin{array}{rcl}             
\X_{w_1}^c   & = &  Z_{w_1} \left(c-\X_{j}^c \right) \\                  
  &\vdots &  \\          
\X_{w_{d-1}}^c  &=&  Z_{w_{d-1}} \left(c-\X_{j}^c\right) \prod_{k=1}^{d-2}\left(1-Z_{w_k} \right) \\
	\end{array} \, .
\end{equation}  

$\quad$ (ii) If $\left(\X_{j}^c, \bo{\X}_{\sim j}^c \right) \stackrel{d}{=}  \left\{\X_i \sim F_i,\, i=j, \, w_1, \ldots, w_{d-1}\, :\, \sum_{i=1}^d  G_i^{-1}\left(F_{i}(\X_i) \right) <  c \right\}$, then  we have  $\X_{j}^c \stackrel{d}{=} F_{j}^{-1}\left(G_{j} (Z_{j})\right)$,
\begin{equation} \label{eq:gacop2}     
\displaystyle 
\begin{array}{rcl}            
\X_{w_1}^c   & = & F_{w_1}^{-1}\left(G_{w_1} \left( Z_{w_1} \left(c-  G_{j}^{-1} \left(F_{j}\left(\X_{j}^c \right)\right) \right) \right) \right) \\
  &\vdots &  \\          
\X_{w_{d-1}}^c  &=&  F_{w_{d-1}}^{-1}\left(G_{w_{d-1}} \left( Z_{w_{d-1}} \left(c-  G_{j}^{-1} \left(F_{j}\left(\X_{j}^c \right)\right)\right) \prod_{k=1}^{d-2}\left(1-Z_{w_k} \right) \right) \right)  \\      
	\end{array} \, . 
\end{equation}   
\end{corollary}        
\begin{preuve}   
See Appendix \ref{app:theo:idconstga2}. 
\begin{flushright}
$\Box$ 
\end{flushright}
\end{preuve}  

Corollaries \ref{coro:idconstga}-\ref{coro:idconstga2} provide DMs for some constrained gamma distributions. As the exponential distribution (i.e., $\Gamma(1,\, \beta)$) and the chi-square $\khi^2(\nu)$ distribution with $\nu$ degrees of freedom (i.e., $\Gamma(\nu/2,\, 1/2)$) are particular cases of the gamma distribution, Corollaries \ref{coro:idconstga}-\ref{coro:idconstga2} can be used to obtain the DMs for constrained exponential distributions and constrained chi-square distributions.
 
%%-------------------------------------
\subsection{Gaussian random variables under constraints and related distributions} 
%------      
This section deals with independent Gaussian variables under linear and quadratic constraints. First, we provide DMs for Gaussian distributions under a linear constraint (see Theorem \ref{theo:idconstg}). Second, we derive DMs of uniformly-distributed variables on a sphere of $\R^d$ (see Corollary \ref{coro:idconstgau2}) and in a sphere of $\R^d$ (see Corollary \ref{coro:idconstgau3}). Dependency models for distributions related to the former distributions are also provided. \\ 

For independent variables $\X_i \sim \mathcal{N}(0,\, \sigma_i^2)$ with $i=1, \ldots, d$ and $j\in \{1, \ldots, d\}$, we use   
$\Sigma^c := \left[\Sigma_{i\ell}^c :=\frac{\sigma^2_{i}}{\sum_{k=1}^d \sigma^2_{k}} \left(\delta_{i \ell} \left(\sigma^2_{\ell} + \sum_{\substack{k=1,\\ k \neq i}}^d \sigma^2_{k} \right)-\sigma^2_{\ell}\right), \: i,\, \ell = j,\, w_1, \ldots, w_{d-2}\right]$ for a symmetric and positive definite matrix that belongs to $ \R^{d-1 \times d-1}$ and $\mathcal{L}$ for its Cholesky factor where $\delta_{i \ell} = 1\, \mbox{if}\; i=\ell \, \mbox{and}\; 0 \, \mbox{otherwise}$. We also use $\Phi$ for the CDF of the standard Gaussian distribution  and $F_i$ with $i=1, \ldots, d$ for continuous CDFs. 
  
%-------------------------                                                     
\begin{theorem} \label{theo:idconstg}  
Let $\bo{Z}_{\sim j} \sim \mathcal{N}_{d-2}\left(\bo{0},\, \mathcal{I}\right)$ and $Z_{j} \sim \mathcal{N}\left(\frac{c\,\sigma^2_{j}}{\sum_{k=1}^d \sigma^2_{k}}, \, \Sigma_{j j}^c \right)$ be independent variables. \\
  
$\quad$ (i) If $\left(\X_{j}^c,\,\bo{\X}^c_{\sim j}\right)  \stackrel{d}{=} \left\{\X_i \sim \mathcal{N}(0,\, \sigma_i^2),\, i=j,\, w_1, \ldots, w_{d-1} \, :\, \sum_{i=1}^d \X_i = c \right\}$, then we have $\X_{j}^c \stackrel{d}{=} Z_{j}$, 
%----- 
\begin{eqnarray} \label{eq:gaucons}        
\displaystyle                                                        
\left[\begin{array}{c}               
\X_{w_1}^c  \\                    
  \vdots   \\          
\X_{w_{d-2}}^c   
	\end{array}\right]
	& \stackrel{\text{}}{=} & \left[ \mathcal{L} \,  
         \left[\begin{array}{c} \frac{\left(\X_{j}^c - \frac{c\,\sigma^2_{j}}{\sum_{k=1}^d \sigma^2_{k}} \right)}{\sqrt{\Sigma_{j j}^c}} \\
	          \bo{Z}_{\sim j}  \\
							\end{array} 
			\right] \right]_{\sim 1}    
	+  \left[\begin{array}{c}              
 \frac{c\,\sigma^2_{w_1}}{\sum_{k=1}^d \sigma^2_{k}}  \\                  
  \vdots   \\          
\frac{c\,\sigma^2_{w_{d-2}}}{\sum_{k=1}^d \sigma^2_{k}} 
	\end{array}\right] \, ,    \\     
\X_{w_{d-1}}^c  &=& c- \X_{j}^c- \sum_{i=1}^{d-2} \X_{w_i}^c  \nonumber \, ,   
\end{eqnarray}
with $[\bu]_{\sim 1}$ a vector obtained by excluding the first element of $[\bu]$.\\ 

$\quad$ (ii) If $\left(\X_{j}^c,\,\bo{\X}^c_{\sim j}\right)  \stackrel{d}{=}  \left\{\X_i \sim F_i,\, i=j,\, w_1, \ldots, w_{d-1} \, :\, \sum_{i=1}^d \sigma_i \Phi^{-1}\left(F_i(\X_i)\right) = c \right\}$, then
$\X_{j}^c \stackrel{d}{=} F_{j}^{-1}\left(\Phi\left(\frac{Z_{j}}{\sigma_{j}} \right) \right)$, 
%-------    
\begin{equation} \label{eq:gauconsco} 
\displaystyle                               
\begin{array}{rcl}       
\X_{w_1}^c   & = &   F_{w_1}^{-1}\left( \Phi\left(\frac{Y_{w_1}}{\sigma_{w_1}} \right) \right)\\  
  &\vdots &  \\          
\X_{w_{d-1}}^c  &=&   F_{w_{d-1}}^{-1}\left(\Phi\left(\frac{Y_{w_{d-1}}}{\sigma_{w_{d-1}}} \right)\right) \\ 
	\end{array} \, ,
\end{equation}      
where 
\begin{eqnarray} 
\displaystyle                                                        
\left[\begin{array}{c}              
Y_{w_1}^c  \\                  
  \vdots   \\          
Y_{w_{d-2}}^c  
	\end{array}\right]
	& \stackrel{\text{}}{=} & \left[ \mathcal{L} \, 
         \left[\begin{array}{c} \frac{\left(\sigma_{j}\Phi^{-1} \left( F_{j}(\X_{j}^c)\right) - \frac{c\,\sigma^2_{j}}{\sum_{k=1}^d \sigma^2_{k}} \right)}{\sqrt{\Sigma_{j j}^c}} \\
	          \bo{Z}_{\sim j}  \\
							\end{array} 
			\right] \right]_{\sim 1}   
	+  \left[\begin{array}{c}              
 \frac{c\,\sigma^2_{w_1}}{\sum_{k=1}^d \sigma^2_{k}}  \\                  
  \vdots   \\          
\frac{c\,\sigma^2_{w_{d-2}}}{\sum_{k=1}^d \sigma^2_{k}} 
	\end{array}\right]     \nonumber \\     
	Y_{w_{d-1}}^c  &=& c- \sigma_{j} \Phi^{-1} \left( F_{j}(\X_{j}^c)\right) - \sum_{i=1}^{d-2} Y_{w_i}^c  \nonumber \,  .   
\end{eqnarray}             
\end{theorem}      
\begin{preuve}   
See Appendix \ref{aap:theo:idconstg}.   
\begin{flushright}
$\Box$   
\end{flushright} 
\end{preuve}     

By combining Theorem \ref{theo:idconstg} with Remark \ref{rem:lcdepm}, we can derive DMs for other distributions.  
For the quadratic constraints such as $\bo{\X}^\T \mathcal{A} \bo{\X} =c$ and $\bo{\X}^\T \mathcal{A} \bo{\X} < c$ with $\mathcal{A}$ a symmetric and positive definite matrix, it is interesting to have the DMs of independent Gaussian variables under such constraints (see Corollaries \ref{coro:idconstgau2}-\ref{coro:idconstgau3}).  Namely, consider a  Rademacher-distributed variable $R$ and the generalized beta distribution of the first-kind, that is, $GB1\left(p, q, a, b \right)$ (\cite{mcdonald95}).
                                 
%--------------------------------------------                  
\begin{corollary} \label{coro:idconstgau2}  
%------
Let $Z_{w_i} \sim GB1\left(2,\, 1,\, \frac{1}{2}, \, \frac{d-i-1}{2}\right),\, i=1, \ldots, d-2$, $Z_{j} \sim GB1\left(2,\, \sqrt{c}, \,\frac{1}{2}, \, \frac{d-1}{2} \right)$ and $R_i \sim R$ with $i=1, \ldots, d$ be independent variables; $F_i$ be a continuous CDF. \\

$\quad$ (i) If $\left(\X_{j}^c,\,\bo{\X}^c_{\sim j}\right)  \stackrel{d}{=}  \left\{\X_i \sim \mathcal{N}(0,\, 1),\,  i=j,\, w_1, \ldots, w_{d-1} \, :\, \sum_{i=1}^d \X_i^2 = c \right)$, then we have   $\X_{j}^c \stackrel{d}{=} R_{j} Z_{j}$,
%------------       
\begin{equation} \label{eq:gausqua}      
\displaystyle                                                        
\begin{array}{rcl}           
\X_{w_1}^c   & = &  R_{w_1} Z_{w_1} \sqrt{c- \left(\X_{j}^c \right)^2} \\
  &\vdots &  \\           
\X_{w_{d-2}}^c  &=&  R_{w_{d-2}} Z_{w_{d-2}} \sqrt{\left(c- \left(\X_{j}^c \right)^2 \right) \prod_{k=1}^{d-3}\left(1-Z_{w_k}^2 \right) } \\
 \X_{w_{d-1}}^c &=&  R_{w_{d-1}} \sqrt{ \left(c- \left(\X_{j}^c \right)^2 \right) \prod_{k=1}^{d-2}\left(1-Z_{w_k}^2 \right)} \\
	\end{array} \, .  
\end{equation} 
   
$\quad$ (ii) If $\left(\X_{j}^c,\,\bo{\X}^c_{\sim j}\right)  \stackrel{d}{=} \left\{\X_i \sim  F_i,\,  i=j,\, w_1, \ldots, w_{d-1}\, :\, \sum_{i=1}^d \left[\Phi^{-1} \left(F_i(\X_i)\right)\right]^2 = c \right\}$, then   $\X_{j}^c \stackrel{}{=} F_{j}^{-1}\left( \Phi\left( R_{j} Z_{j} \right)\right)$,  
%-------
\begin{equation} \label{eq:gausquaco}
\displaystyle  
\begin{array}{rcl}
\X_{w_1}^c   & = & F_{w_1}^{-1}\left(\Phi \left(R_{w_1} Z_{w_1} \sqrt{c- \left(\Phi^{-1} \left(F_{j}(\X_{j}^c)\right)  \right)^2 }\right) \right) \\
  &\vdots &  \\          
\X_{w_{d-2}}^c  &=&  F_{w_{d-2}}^{-1}\left(\Phi \left(R_{w_{d-2}}Z_{w_{d-2}} \sqrt{\left(c- \left(\Phi^{-1} \left(F_{j}(\X_{j}^c)\right)  \right)^2 \right) \prod_{k=1}^{d-3}\left(1-Z_{w_k}^2 \right) } \right) \right)  \\
 \X_{w_{d-1}}^c &=&  F_{w_{d-1}}^{-1}\left(\Phi \left(R_{w_{d-1}} \sqrt{ \left(c- \left(\Phi^{-1} \left(F_{j}(\X_{j}^c)\right)  \right)^2 \right) \prod_{k=1}^{d-2}\left(1-Z_{w_k}^2 \right)} \right) \right)  \\ 
	\end{array} \, .  
\end{equation}      
\end{corollary}   
\begin{preuve}  
See Appendix \ref{app:coro:idconstgau2}. 
\begin{flushright}   
$\Box$   
\end{flushright} 
\end{preuve} 
     
When $c=1$, Equation (\ref{eq:gausqua}) is distributed as $\bo{\X}_c' :=\left(\frac{X_1^2}{\sum_{i=1}^d X_i^2}, \, \ldots,\, \frac{\X_d^2}{\sum_{i=1}^d \X_i^2} \right)$ where $\X_j \sim \mathcal{N}(0,\,1)$, $j=1, \ldots, d$. It is to be noted that Equation (\ref{eq:gausqua}) can be used to generate random values that are uniformly distributed on the unit sphere in $\R^d$ (\cite{devroye86}). Using Equations (\ref{eq:gausqua})-(\ref{eq:gausquaco}), we are also able to provide other DMs. For instance, we can model
$ 
\bo{Y}^c \stackrel{d}{=} \left\{\bo{\X} \sim \mathcal{N}(\bo{0},\, \Sigma)\, :\, \bo{\X}^\T \Sigma^{-1} \bo{\X} =c \right\}$ as follows: $\bo{Y}^c \stackrel{}{=} \mathcal{L} \bo{\X}^c $  
with $\bo{\X}^c$ given by (\ref{eq:gausqua}) and $\mathcal{L}$ the Cholesky factor of $\Sigma$. Thus, we have  $Y_{j}^c \stackrel{}{=} R_{j} V$ with $V \sim GB1\left(2,\, \sqrt{c \Sigma_{j j}}, \,\frac{1}{2}, \, \frac{d-1}{2} \right)$ and 

$$      
\displaystyle                                                          
\left[\begin{array}{c}     
Y_j^c   \\  
Y_{w_1}^c   \\      
\vdots   \\       
Y_{w_{d-2}}^c  \\
 Y_{w_{d-1}}^c \\
	\end{array}\right]
	\stackrel{}{=}  \mathcal{L} \,  
	\left[	\begin{array}{c}     
  \frac{Y_{j}^c}{\sqrt{\Sigma_{j j}}} \\
  R_{w_1} Z_{w_1} \sqrt{c- \frac{\left(Y_{j}^c\right)^2}{\Sigma_{j j}}} \\
  \vdots   \\           
  R_{w_{d-2}} Z_{w_{d-2}} \sqrt{ \left(c- \frac{\left(Y_{j}^c\right)^2}{\Sigma_{j j}} \right) \prod_{k=1}^{d-3}\left(1-Z_{w_k}^2 \right) } \\
  R_{w_{d-1}} \sqrt{ \left(c- \frac{\left(Y_{j}^c\right)^2}{\Sigma_{j j}} \right) \prod_{k=1}^{d-2}\left(1-Z_{w_k}^2 \right)}\\
	\end{array} \right] 	\, . 
	$$      
    
Likewise, a DM for a random vector uniformly distributed in the unit sphere in $\R^d$ can be obtained using Corollary \ref{coro:idconstgau3}.  
    
%-------------------------                                          
\begin{corollary} \label{coro:idconstgau3}  
%------
Let $Z_{w_i} \sim GB1\left(2,\, 1,\,\frac{1}{2}, \, \frac{d-i}{2}\right),\, i=1, \ldots, d-1$, $Z_{j} \sim GB1 \left(2,\, \sqrt{c},\, \frac{1}{2}, \, \frac{d}{2} \right)$ and $R_j \sim R$ with $i=1, \ldots, d$ be independent variables; $F_i$ be a continuous CDF. \\
    
$\quad$ (i) If $\left(\X_{j}^c,\,\bo{\X}^c_{\sim j}\right)  \stackrel{d}{=} \left\{\X_i \sim \mathcal{N}(0,\, 1),\,  i=j,\, w_1, \ldots, w_{d-1} \, :\, \sum_{i=1}^d \X_i^2 < c \right)$, then we have  $\X_{j}^c \stackrel{d}{=} R_{j}  Z_{j}$,
%------------    
\begin{equation} \label{eq:gausqua1}     
\displaystyle                                                        
\begin{array}{rcl}           
\X_{w_1}^c   & = &  R_{w_1}  Z_{w_1} \sqrt{c- \left(\X_{j}^c \right)^2} \\
  &\vdots &  \\           
\X_{w_{d-1}}^c  &=&  R_{w_{d-1}} Z_{w_{d-1}} \sqrt{ \left(c- \left(\X_{j}^c \right)^2 \right) \prod_{k=1}^{d-2}\left(1-Z_{w_k}^2 \right) } \\
	\end{array} \, .   
\end{equation}

$\quad$ (ii) If $\left(\X_{j}^c,\,\bo{\X}^c_{\sim j}\right)  \stackrel{d}{=} \left\{\X_i \sim  F_i,\,  i=j,\, w_1, \ldots, w_{d-1} \, :\, \sum_{i=1}^d \left[\Phi^{-1} \left(F_i(\X_i)\right)\right]^2 < c \right)$, then 
$\X_{j}^c \stackrel{d}{=} F_{j}^{-1}\left( \Phi\left( R_{j} Z_{j} \right) \right)$, 
%-------
\begin{equation} \label{eq:gausquaco1}    
\displaystyle 
\begin{array}{rcl} 
\X_{w_1}^c   & = & F_{w_1}^{-1}\left(\Phi \left(R_{w_1} Z_{w_1} \sqrt{c- \left(\Phi^{-1} \left(F_{j}(\X_{j}^c)\right)  \right)^2}\right) \right) \\
  &\vdots &  \\          
\X_{w_{d-1}}^c  &=&  F_{w_{d-1}}^{-1}\left(\Phi \left(R_{w_{d-1}} Z_{w_{d-1}} \sqrt{ \left(c- \left(\Phi^{-1} \left(F_{j}(\X_{j}^c)\right)  \right)^2 \right) \prod_{k=1}^{d-2}\left(1-Z_{w_k}^2 \right) } \right) \right)  \\  
	\end{array} \, .
\end{equation}
\end{corollary}     
\begin{preuve} 
The proofs of Points (i) -(ii) are similar to those of Corollaries \ref{coro:idconstga2}-\ref{coro:idconstgau2}. 
\begin{flushright}  
$\Box$     
\end{flushright}   
\end{preuve}

%%%%%%%%%%%%%%%%%%%%%%%%%%%%%%%%%%%%%%%%%%%% 
%------------------------	
\section{Sensitivity-based approach for choosing the efficient dependency model} \label{sec:stra}
%--- 
It comes out from Section \ref{sec:pdm} that there is $d$ possible DMs for a given $d$-dimensional random vector, regarding the choice of $\X_{j}$ as input with $j \in \{1,\,\ldots,\, d\}$. Regarding the outputs $\bo{\X}_{\sim j}$, we have $(d-1)!$ possibilities. We then have in total $d!$ possibilities when no supplementary information such as the causality is available. This section shows how multivariate sensitivity analysis (\cite{lamboni11,gamboa14,lamboni18a,lamboni19}), including Sobol' indices allows for identifying the efficient DM.
            
%-----------------------------------   
\subsection{Multivariate sensitivity analysis: generalized sensitivity indices}
%---    
MSA allows for identifying the input variables of a given model $\M : \R^d \to \R^\NN$ that significantly contribute to the model variance-covariance. It is based on sensitivity functionals (SFs), which contain the information about the single and overall contributions of an input or a group of inputs over the whole model outputs. \\
 
 Namely, let us recall that for $u \subseteq \{1,\, \ldots,\, d\}$, $\bo{\X}_u =(\X_j,\, j\in u)$ and $\bo{\X}_{\sim u} := (\X_j,\, j \in \{1,\, \ldots,\, d\}\setminus u)$. While the first-order SF, that is,  
$$
\M_u^{fo}(\bo{\X}_u):=\esp \left[\M(\bo{\X})| \bo{\X}_{u}\right] -\esp \left[\M(\bo{\X})
\right] \, ,
$$    
is used to assess the single contribution of the inputs $\bo{\X}_u$ over the whole output(s), the total SF given by
$$
\M^{tot}_u(\bo{\X}):= \M(\bo{\X}) - \esp \left[\M(\bo{\X})| \bo{\X}_{\sim u}\right] \, , 
$$ 
is used to measure the overall contribution of $\bo{\X}_u$, including interactions (\cite{lamboni18,lamboni18a,lamboni19}). The generalized sensitivity indices (GSIs) from MSA are given below.
               
\begin{defi}            \label{def:gsi} 
Let $\bo{\X}$ be independent variables; $\Sigma := \var\left[\M(\bo{\X})\right]$, $\D_u := \var\left[\M_u^{fo}(\bo{\X}_u)\right]$, and $\D_u^{tot} := \var\left[\M_u^{tot}(\bo{\X})\right]$ be the variances of the model outputs, the first-order and total SFs,  respectively and $\normf{\Sigma}^2 = \trace\left(\Sigma \Sigma^\T \right)$. \\   
  
$\quad$ (i) The first-type GSIs are defined below (\cite{lamboni11,gamboa14,lamboni18a,lamboni19}). \\  
%-  
The first-order GSI of $\bo{X}_u$ is given by 
\begin{equation}        \label{eq:gsif}   
GSI_u := \frac{\trace(\D_u)}{\trace(\Sigma)} \, .
\end{equation}     
Further, the total GSI of $\bo{X}_u$ is given by 
\begin{equation}  \label{eq:gsitf} 
GSI_{T_u} := \frac{\trace\left(\D_u^{tot}\right)}{\trace(\Sigma)} \, . 
\end{equation}
         
$\quad$ (ii) The second-type GSIs are defined as follows (\cite{lamboni18a,lamboni19}):  
%-    
\begin{equation}        \label{eq:gsil2}      
GSI_u^{F} := \frac{\normf{\D_u}}{\normf{\Sigma}} \, ; 
\end{equation}      
\begin{equation}  \label{eq:gsitl2} 
GSI_{T_u}^{F} := \frac{\normf{\D_u^{tot}}}{\normf{\Sigma}} \, . 
\end{equation}
\end{defi}        

It is worth noting that the total index measures the overall contribution of inputs, including interactions. It is used to select the input variables that lead to the highest reduction of the model variance-covariance once such variables are known.
                                       
\begin{rem}   \label{rem:sob}
For a single output ($\NN=1$), both types of GSIs come down to Sobol' indices (\cite{sobol93,saltelli00}). The second-type GSIs account for the correlations among SFs. 
\end{rem}         
                        
%-----------------------------------
\subsection{Choice of the efficient dependency model}
%----             							     
For a $d$-dimensional random vector $\bo{\X}$, a DM of $\bo{\X}$ is given by Equation (\ref{eq:depmpr}), that is,     
\begin{equation} 
\bo{\X}_{\sim j} = r_{j}\left(\X_{j}, \bo{Z}_{\sim j} \right) \, , \nonumber
\end{equation}  
when $\X_{j}$ is used as input for all $j \in\{1, \ldots, d\}$. As the function $r_{j}$ includes only independent variables, we can use Definition  \ref{def:gsi} to quantify the importance of $\X_{j}$. The input variables $\bo{Z}_{\sim j}$ can be seen as auxiliary variables, as such variables may be un-meaningful regarding the phenomena of interest. In (\ref{eq:depmpr}), $\bo{\X}_{\sim j}$ is the output(s), and we have $(d-1)!$ possible choices of the outputs $\bo{\X}_{\sim j}$ due to the permutation of $\{1, \ldots, d\} \setminus\{j\}$. For two different permutations $w,\, v$ of $\{1, \ldots, d\} \setminus\{j\}$, we see that 
$$ 
\normf{\var(\bo{\X}_{w})} = \normf{\var \left(\bo{\X}_{v}\right)}, 
\qquad 
\trace(\var(\bo{\X}_{w})) = \trace(\var \left(\bo{\X}_{v}\right)) \, .
$$ 
 Since both choices of outputs have the same measure of variability used to define GSIs (see Definition \ref{def:gsi}), the permutation of the outputs $\bo{\X}_{\sim j}$ does not affect the GSIs of $\X_j$. Therefore, the $(d-1)!$ possibilities do not matter regarding the importance measure used in this paper. We then have only $d$ potential DMs.\\

In mathematical and statistical modeling, a model output with the smallest variance is qualified as the best model. The variance of the DM output(s) $\bo{\X}_{\sim j}$ represents the variance of $\bo{\X}$ that remains once $\X_j$ is used as input. The higher is the variance of $\X_j$, the smaller is the output(s) variance. Since the $d$ DM output(s) are different, the ratio between the variance of $\X_j$ and the output(s) variance given by GSIs (see Definition \ref{def:gsi}) serves as an optimal criterion for comparing the $d$ models. Indeed, the higher is the total GSI, the better is the model independently of the variances of the $d$ output(s) because 
$
 \left(\X_j, r_{j}\left(\X_{j}, \bo{Z}_{\sim j} \right)\right), j=1, \ldots, d
$ have the same values of $\normf{\var(\bo{\X})}$ and $\trace[\var(\bo{\X})]$. Thus, Theorem \ref{theo:ch} provides the efficient DM.
         
%-----------------------                                                   
\begin{theorem}    \label{theo:ch} 
Let $GSI_{T_{j}}^{F}$ be the second-type total GSI of $\X_{j}$ associated with the DM $r_j\left(\X_{j},\bo{Z}_{\sim j} \right)$ with $j =1, \ldots, d$, and assume  $\esp\left[\norme{\bo{\X}}^2 \right] < +\infty$. \\    
   
 Then,  the efficient DM is  given by
%----
\begin{equation} 
\bo{\X}_{\sim j^*} = r_{j^*}\left(\X_{j^*}, \bo{Z}_{\sim j^*} \right) \, ,
\end{equation} 
with
$$
j^* =\arg\!\max_{ j \in\{1,\ldots, d\}} GSI_{T_{j}}^{F} \, .
$$
\end{theorem}   
\begin{preuve} 
 First, all the $d$ models are completely determined by the choice of the input $\X_{j}$. 
 Second, if we use $GSI_{T_{j*}}^{F} := \max \left( GSI_{T_{1}}^{F} , \ldots, GSI_{T_{d}}^{F} \right)$ for the highest value of the total GSIs, it follows that $\X_{j*}$ has the highest contribution regarding the variability of the $d$ DMs.  
\begin{flushright}    
$\Box$    
\end{flushright} 
\end{preuve}

In Theorem \ref{theo:ch}, we use the total GSIs of the second-type to identify $j^*$, as such indices account for the correlations among SFs.  When $j^*$ is not unique, we have to use the first-type total indices (i.e., $GSI_{T_{j}}$) in order to expect finding $j^*$. In the same sense, if the latter indices are not helpful, we have to rely on the first-order GSIs for choosing the efficient DM. However, in presence of exchangeable CDF of $\bo{\X}$, which remains unchanged when permuting its components, we should have the same indices for the $d$ models. Thus, the $d$ models are equivalent and can be used. 
                  
%-------------------------------------            
%------------------------------------            
\subsection{Test cases}\label{sec:num} 
%-----	    
In this section, we consider two random vectors of dependent variables to illustrate our approach. For each random vector, we provide the efficient DMs according to the distributions parameters.
 
%-------------------------                           
\subsubsection{Gaussian distribution ($d=3$)}    
%-----   
 We consider three correlated variables given by  
$$ 
\bo{\X} \sim \mathcal{N}\left(\bo{0},\, \left[\begin{array}{ccc}
\sigma_1^2 &\rho_{12} \sigma_1\sigma_2 & \rho_{13}\sigma_1\sigma_3 \\ 
\rho_{12}\sigma_1\sigma_2 & \sigma_2^2 & \rho_{23}\sigma_2\sigma_3  \\
\rho_{13}\sigma_1\sigma_3 & \rho_{23}\sigma_2\sigma_3  & \sigma_{3}^2    
\end{array}\right]\right) \, . 
$$                        

The three DMs of $\bo{\X}$ are given by (see \cite{lamboni21}) 
$$     
r_1 (\X_1,\, Z_2,\, Z_3) =
 \left[\begin{array}{c}   
       \frac{\rho_{12}\sigma_2}{\sigma_1}\X_1 + \sqrt{1-\rho_{12}^2}  \Z_2   \\
			  \frac{\rho_{13}\sigma_3}{\sigma_1}\X_1 + \frac{\sigma_3(\rho_{23}-\rho_{12}\rho_{13})}{\sigma_2 \sqrt{1-\rho_{12}^2}} \Z_2 + \sqrt{\frac{1-\rho_{12}^2-\rho_{13}^2-\rho_{23}^2+2\rho_{12}\rho_{13}\rho_{23}}{1-\rho_{12}^2}}    \Z_3        
      \end{array} \right]   \, ,                      
$$           
$$
r_2 (Z_1,\, \X_2,\, Z_3) =
 \left[\begin{array}{c}            
       \frac{\rho_{12}\sigma_1}{\sigma_2}\X_2 + \sqrt{1-\rho_{12}^2}  \Z_1   \\
	 \frac{\rho_{23}\sigma_3}{\sigma_2}\X_2 + \frac{\sigma_3(\rho_{13}-\rho_{12}\rho_{23})}{\sigma_1 \sqrt{1-\rho_{12}^2}} \Z_1 + \sqrt{\frac{1-\rho_{12}^2-\rho_{13}^2-\rho_{23}^2+2\rho_{12}\rho_{13}\rho_{23}}{1-\rho_{12}^2}}    \Z_3              
      \end{array} \right]  \, ,                       
$$
$$ 
r_3(Z_1,\, Z_2,\, \X_3) =   
 \left[\begin{array}{c}       
     \frac{\rho_{13}\sigma_1}{\sigma_3}\X_3 + \sqrt{1-\rho_{13}^2}  \Z_1   \\
	 \frac{\rho_{23}\sigma_2}{\sigma_3}\X_3 + \frac{\sigma_2(\rho_{12}-\rho_{13}\rho_{23})}{\sigma_1 \sqrt{1-\rho_{13}^2}} \Z_1 + \sqrt{\frac{1-\rho_{12}^2-\rho_{13}^2-\rho_{23}^2+2\rho_{12}\rho_{13}\rho_{23}}{1-\rho_{13}^2}}    \Z_2 \\
      \end{array} \right]  \, ,     
$$
where $\Z_j \sim \mathcal{N}(0,\, \sigma_j^2)$ with $j=1,\, 2,\,3$ are independent.\\

Using Definition  \ref{def:gsi}, the total-effect variance of $\X_j$, that is, $\Sigma_j^{tot}$ associated with the three DMs are given by 
$$
\Sigma_1^{tot} :=  
 \left[\begin{array}{cc}       
		 \rho_{12}^2\sigma_2^2 &  \rho_{12} \rho_{13}\sigma_2 \sigma_3 \\
	  \rho_{12} \rho_{13}\sigma_2 \sigma_3 & \rho_{13}^2\sigma_3^2
      \end{array} \right]  \, ,  
			\quad
\Sigma_2^{tot} :=    
 \left[\begin{array}{cc}       
   \rho_{12}^2 \sigma_1^2 &  \rho_{12}\rho_{23}\sigma_1\sigma_3 \\
		\rho_{12}\rho_{23}\sigma_1\sigma_3 & \rho_{23}^2\sigma_3^2
      \end{array} \right]  \, ,                                
$$     
$$      
\Sigma_3^{tot} :=      
 \left[\begin{array}{cc}       
   \rho_{13}^2 \sigma_1^2 & \rho_{13} \rho_{23} \sigma_1\sigma_2  \\
		\rho_{13} \rho_{23} \sigma_1\sigma_2 & \rho_{23}^2 \sigma_2^2 \\
      \end{array} \right]  \, ,                              
$$      
and the total generalized sensitivity indices are given by

$$
GSI_{T_1}^F :=\left(\frac{\rho_{12}^4\sigma_2^4 + \rho_{13}^4\sigma_3^4 +2 \rho_{12}^2 \rho_{13}^2\sigma_2^2 \sigma_3^2}{\sigma_2^4 + \sigma_3^4 +2 \rho_{23}^2 \sigma_2^2\sigma_3^2} \right)^{1/2}\, , 
$$
$$
GSI_{T_2}^F :=\left(\frac{\rho_{12}^4\sigma_1^4 + \rho_{23}^4\sigma_3^4 +2 \rho_{12}^2 \rho_{23}^2\sigma_1^2 \sigma_3^2}{\sigma_1^4 + \sigma_3^4 +2 \rho_{13}^2 \sigma_1^2\sigma_3^2} \right)^{1/2}  \, , 
$$        
$$ 
GSI_{T_3}^F :=\left(\frac{\rho_{13}^4\sigma_1^4 + \rho_{23}^4\sigma_2^4 +2 \rho_{13}^2 \rho_{23}^2\sigma_1^2 \sigma_2^2}{\sigma_1^4 + \sigma_2^4 +2 \rho_{12}^2 \sigma_1^2\sigma_2^2} \right)^{1/2}  \, .   
$$  
    
\begin{figure}[!hbp] 
\begin{center}
\includegraphics[height=10cm,width=15.5cm,angle=0]{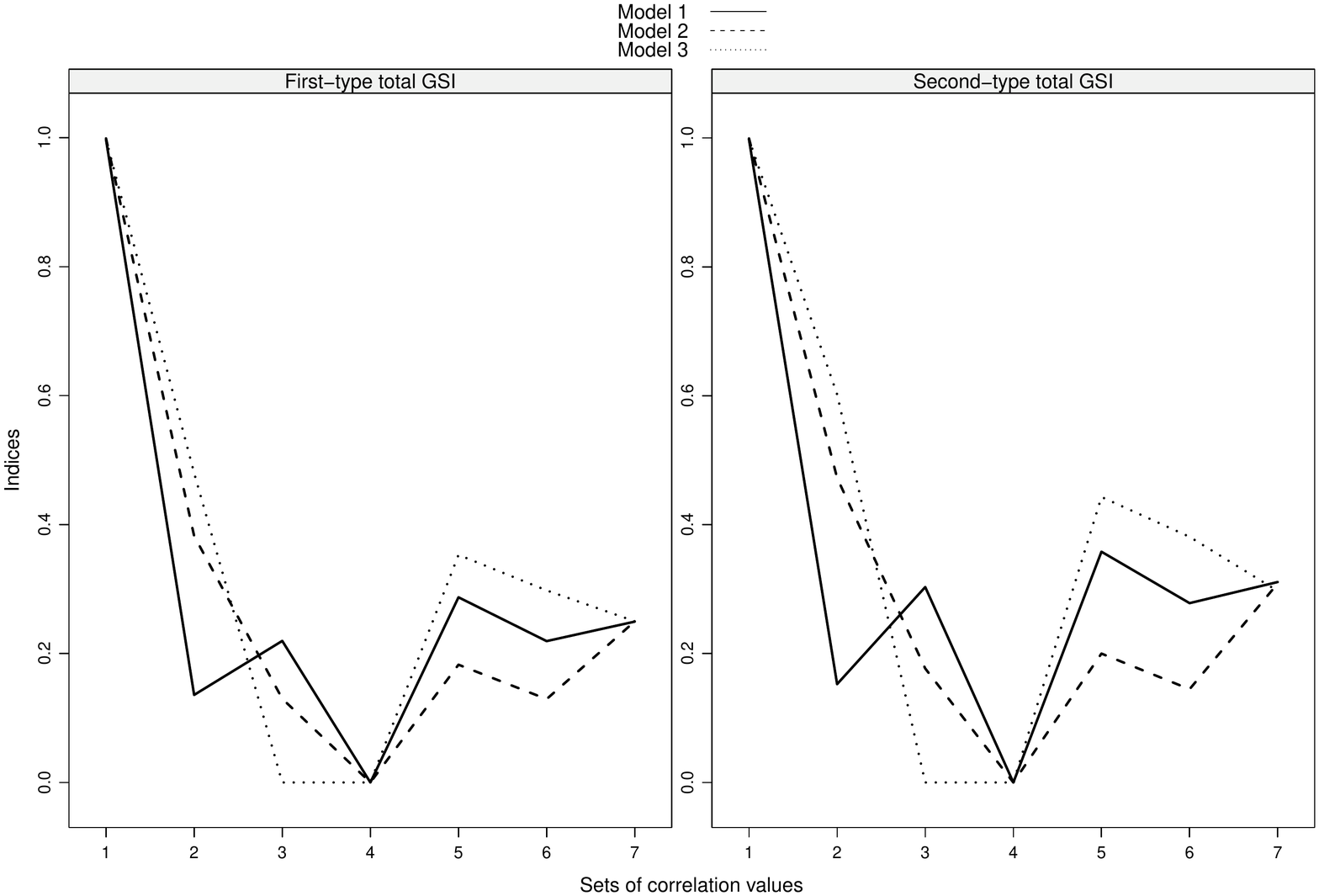}
\end{center}
\caption{Total generalized sensitivity indices of the first-type and second-type for the three dependency models with respect to the sets of correlations listed in Table \ref{tab:norm}.}   
 \label{fig:nor}               
\end{figure}

Figure \ref{fig:nor} shows the importance of the three DMs  for  different values of $\rho_{12}, \, \rho_{13},\, \rho_{23}$ listed in Table \ref{tab:norm} and for $\sigma_1 =3,\, \sigma_2=5, \, \sigma_3 =4$. Regarding the second-type total GSIs of the three models (see Theorem \ref{theo:ch}), the model $r_3$ is the best one for the sets of correlations $S_2,\,S_5, \,S_6$, and $r_1$ is the best model for $S_3$. Nevertheless, the three models have the same importance for $S_1,\, S_4,\, S_7$. The set $S_1$ corresponds to almost perfect correlated variables, and $S_4$ corresponds to independent variables. For the set $S_7$, the first-type total GSI does not help identifying the best model as well as the first-order GSIs of both types, showing that the three models are equivalent.      
                  
\begin{table}[ht]   
\centering     
\begin{tabular}{lccc}
  \hline  
	 \hline
Sets & $\rho_{12}$  & $\rho_{13}$ & $\rho_{23}$ \\ 
  \hline
$S_1$ &-0.9990 & 0.9990 & -0.9990 \\  
$S_2$ &  0.2500 & 0.5000 & 0.7500 \\ 
$S_3$ &  0.6000 & 0.0000 & 0.0000 \\  
$S_4$ & 0.0000 & 0.0000 & 0.0000 \\ 
$S_5$ & 0.2500 & 0.8000 & 0.5000 \\ 
$S_6$ & 0.0000 & 0.7500 & 0.4500 \\  
$S_7$ & -0.5000 & 0.5000 & -0.5000 \\     
   \hline            
	 \hline 
\end{tabular}
\caption{Sets of correlation values used for performing multivariate sensitivity analysis.} 
\label{tab:norm}
\end{table}

%-----------------------  
     
%------------------------------------------               
\subsubsection{Trapezoidal domain: ($d=2$)}    
%-----   
 We consider the variables $\bo{\X} \sim \mathcal{U}[0,\; 1]^2$ under the constraint $1-\beta \X_1 -\X_2 \geq 0$ with $\beta \in (0,\; 1]$. The density of $\bo{\X}$ is given by    
$
\rho \left(\bo{\x}\right) =\frac{2}{2-\beta} \indic_{[1-\beta \x_1 -\x_2 \geq 0]}\left(\bo{\x} \right) \indic_{[0,\, 1]^2}\left(\bo{\x} \right)
$.  
                       
We can see that the marginal density of $\bo{\X}$ are given by   
$$   
\rho_1\left(\x_1\right) =  \frac{2}{2-\beta }\left(1- \beta \x_1\right)\indic_{[0,\; 1]}(\x_1)   \, ,   
$$                       
$$
\rho_2\left(\x_2\right) =  \frac{2}{\beta (2-\beta )}\left(1- \x_2\right)\indic_{]1-\beta,\; 1]}(\x_2) + \frac{2}{(2-\beta )}\indic_{[0,\; 1-\beta]}(\x_2) \,
. 
 $$          
Thus, $\X_1$ follows a truncated beta distribution of the first-kind on $[0,\, 1]$, that is,  $\X_1 \sim~B1\left(1/\beta,\, 1,\, 2\right)$ with $[0,\, 1]$ as its support, and $\X_2$ follows the trapezoidal distribution, that is, $\X_2 \sim T\left(0,\, 0,\, 1-\beta,\, 1\right)$. Using Equation (\ref{eq:depmpr}) and $Z_i \sim \mathcal{U}[0,\, 1],\, i=1,\, 2$, the two DMs of $\bo{\X}$ are given by (\cite{lamboni21})
$$   
\X_2= r_1 (\X_1,\,Z_2) =  
    \Z_2\left(1-\beta  \X_1 \right), 
\qquad 
\X_1= r_2 (Z_1,\,\X_2)= Z_1 \min\left(\frac{1-X_2}{\beta },\, 1\right) \, .  
$$

\begin{figure}[!hbp] 
\begin{center}
\includegraphics[height=10cm,width=15.5cm,angle=0]{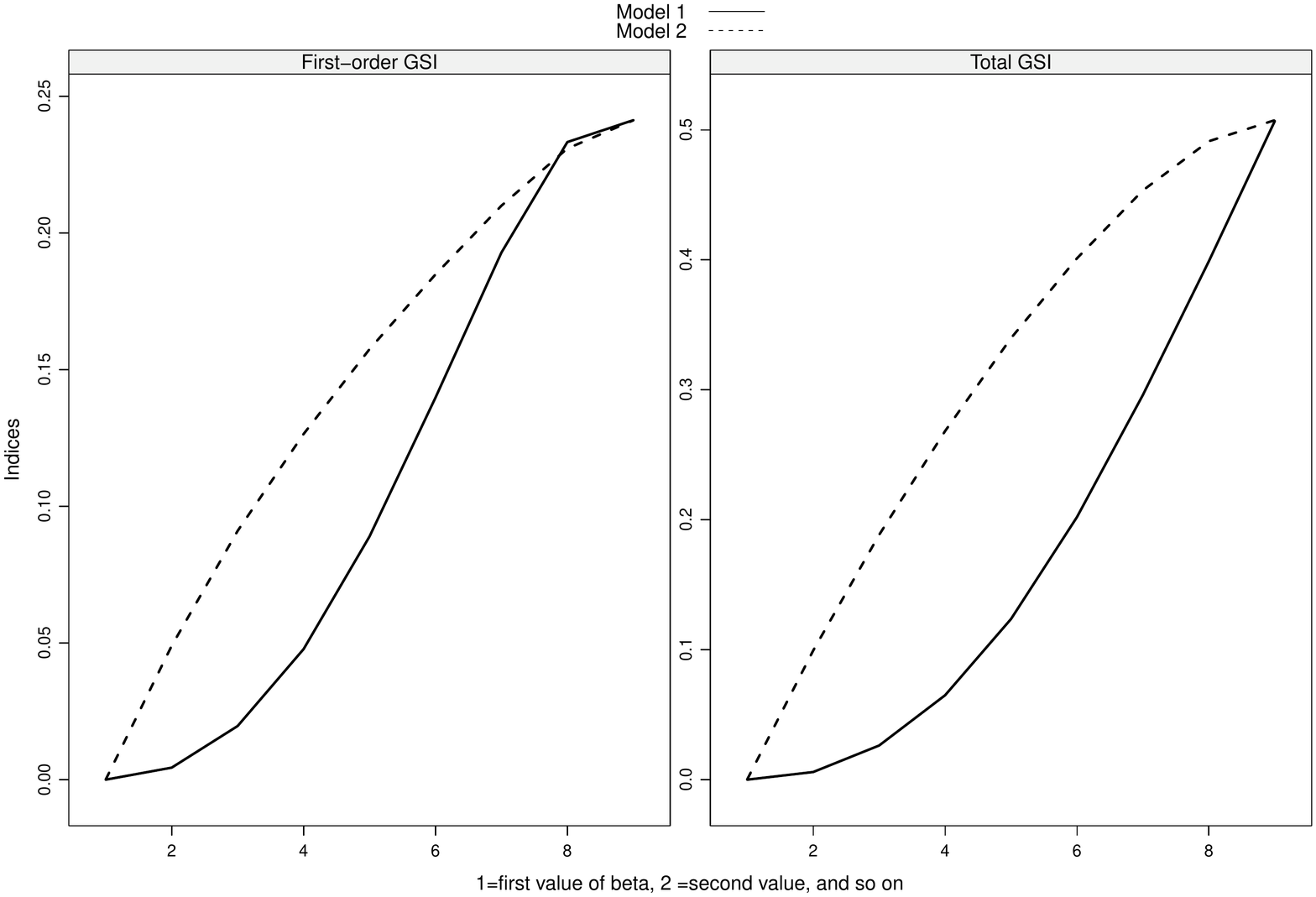}
\end{center}
\caption{First-order and total generalized sensitivity indices of the second-type for the two models with respect to different values of $\beta$.}       
 \label{fig:tra} 
\end{figure}

Since we have only one output for each model, the two types GSIs are equal (see Remark \ref{rem:sob}). 
Figure \ref{fig:tra} shows the importance of both models for different values of\\
 $\beta =\{0.0001, \, 0.125, \, 0.250, \, 0.375, \, 0.500, \, 0.625, \, 0.750, \, 0.875, \, 1\}$. For the first and last values of $\beta$, both models have the same importance. Indeed, the value $\beta=~0$ leads to two independent variables following the standard uniform distribution, and it is obvious that such models have the same importance. When $\beta=1$, the trapezoidal domain comes down to a triangular domain, and both variables have an exchangeable CDF. For other values of $\beta$, we see that the model $r_2$ is the best one. 
  
%-----------------------   
                   
%-----------------------------------------------------                 
%-----------------------------------------------------   
\section{Conclusion} \label{sec:con}  
%-- 
In this paper, we have provided practical dependency models or functions of any $d$-dimensional random vector following classical multivariate distributions such as elliptical distributions for correlated variables, Dirichlet distributions,  independent uniform variables under constraints, independent gamma-distributed variables under constraints, independent Gaussian variables under constraints and related distributions. Our dependency functions are ready to be used for performing uncertainty quantification and sensitivity analysis for complex mathematical models with dependent variables and/or correlated variables introduced in \cite{lamboni21}. \\    
         
Using such dependency models, we are also able to measure the association between the components of a given random vector using multivariate sensitivity analysis. We then provided a methodology for identifying the efficient dependency model, as we have $d$ dependency models for any $d$-dimensional random vector. For exchangeable distributions, the $d$ models are equivalent. Moreover, numerical simulations have shown that the efficient dependency model depends on the parameters of the distribution considered, and when a random vector is consisted of independent variables, the $d$ models collapse, but have the same generalized sensitivity indices. In next future, it is interesting to investigate the impact of sensitivity-based approach for choosing the efficient dependency model on the quality of the design generated by such model.  

%-----------------------------------------------------------------
%----------------------------------------------------------------     
%\section*{Acknowledgments}
% We would like to acknowledge the referees for their comments and suggestions that have helped improving this paper.  
                            
%---------------------------------------------------
%%%%%%%%%%%%%%%%%%%%%%%%%%%%%%%%%%%     
\begin{appendices} 
 %--    
%-----------------------------------  
%-----------------------------------
\section{Proof of Proposition \ref{prop:stco}} \label{app:theo:stco}
%-----  
Since  $\bo{Y} \sim \bo{t}_d(\nu, \boldsymbol{0}, \mathcal{I}) \Longleftrightarrow \bo{\X} \stackrel{d}{=}\mathcal{L}\bo{Y} + \boldsymbol{\mu} \sim \bo{t}_d(\nu, \boldsymbol{\mu}, \Sigma)$ with $\mathcal{L}$ the Cholesky factor of $\Sigma$, we have to show that for all $j_0 \in\{1,\ldots, d\}$
$$ 
\displaystyle     
\left[\begin{array}{c}    
Y_{j_0} \\  
Y_{w_1}= \sqrt{\frac{\nu + Y_{j_0}^2}{\nu +1}} Z_{w_1}  \\         
  \vdots  \\            
 Y_{w_{d-1}}=\sqrt{\frac{\left(\nu + Y_{j_0}^2\right) \prod_{k=1}^{d-2}\left(\nu + k+ (Z_{w_k})^2 \right)}{\prod_{k=1}^{d-1}(\nu +k)}} Z_{w_{d-1}} \\                 
							\end{array}\right] 
											     	\sim \bo{t}_d(\nu, \boldsymbol{0}, \mathcal{I}) \,  .
$$   
For a multivariate $t$-distribution $\bo{Y} \sim \bo{t}_d(\nu, \boldsymbol{0}, \mathcal{I})$, it is knwon that $Y_{w_{k-1}} |\, y_{j_0}, y_{w_1}, \ldots, y_{w_{k-2}} \sim t\left(\nu +k-1, 0, \frac{\nu + y_{j_0}^2+ \sum_{i=1}^{k-2} y_{w_i}^2}{\nu+k-1}\right)$ (see \cite{kotz04} for more details). Thus, it is sufficient to show that $Y_{w_{k-1}} |\, y_{j_0}, z_{w_1}, \ldots, z_{w_{k-2}} \sim t\left(\nu +k-1, 0, \frac{\nu + y_{j_0}^2+ \sum_{i=1}^{k-2} y_{w_i}^2}{\nu+k-1}\right)$ with $k=2,\ldots, d$. The recurrence reasoning leads to the following steps. \\    
 First, we can see that $Y_{w_1} | y_{j_0} \sim t\left(\nu +1, 0, \frac{\nu + y_{j_0}^2}{\nu+1}\right)$, as $Z_{w_1} \sim t\left(\nu +1, 0, 1\right)$.\\
 Second, in one hand, we suppose that $Y_{w_{d-2}} |\, y_{j_0}, z_{w_1}, \ldots, z_{w_{d-3}}  \sim  t \left(\nu +d-2, 0, \frac{\nu + y_{j_0}^2+ \sum_{i=1}^{d-3} y_{w_i}^2}{\nu+d-2} \right)$, and we can write  
\begin{eqnarray}
Y_{w_{d-2}} |\, y_{j_0}, z_{w_1}, \ldots, z_{w_{d-3}}  & \stackrel{d}{=}  & \sqrt{\frac{\nu + y_{j_0}^2+ \sum_{i=1}^{d-3} y_{w_i}^2}{\nu+d-2}} Z_{w_{d-2}} \, , \nonumber 
\end{eqnarray}
as $Z_{w_{d-2}} \sim t\left(\nu + d-2, 0, 1\right)$. The above equation means that for any realization $z_{w_{d-2}}$ of $Z_{w_{d-2}}$, there exists a realization $y_{w_{d-2}}$ of $Y_{w_{d-2}}$ such that 
\begin{equation} \label{eq:eualrea}
y_{w_{d-2}} = \sqrt{\frac{\nu + y_{j_0}^2+ \sum_{i=1}^{d-3} y_{w_i}^2}{\nu+d-2}} z_{w_{d-2}} \Longrightarrow   y_{w_{d-2}}^2 = \frac{\nu + y_{j_0}^2+ \sum_{i=1}^{d-3} y_{w_i}^2}{\nu+d-2} (z_{w_{d-2}})^2  \, .   
\end{equation}   
In the other hand, it comes from the DM that 
\begin{eqnarray}  
Y_{w_{d-2}} |\, y_{j_0}, z_{w_1}, \ldots, z_{w_{d-3}} & \stackrel{d}{=} &
\sqrt{ \frac{\left(\nu + y_{j_0}^2\right) \prod_{i=1}^{d-3}\left(\nu + i+ (z_{w_i})^2 \right)}{\prod_{i=1}^{d-2}(\nu +i)}} Z_{w_{d-2}} \nonumber  \, .  \nonumber
\end{eqnarray} 
 Therefore, we have the following equality   
\begin{equation} \label{eq:eualrea22}
  \frac{\left(\nu + y_{j_0}^2\right) \prod_{i=1}^{d-3}\left(\nu + i+ z_{w_i}^2 \right)}{\prod_{i=1}^{d-2}(\nu +i)} 
=\frac{\nu + y_{j_0}^2+ \sum_{i=1}^{d-3} y_{w_i}^2}{\nu+ d-2} \, . 
\end{equation}
%-- 
Third, as $Z_{w_{d-1}} \sim t\left(\nu +d-1, 0, 1\right)$, we have 
\begin{eqnarray}   
Y_{w_{d-1}} |\, y_{j_0}, z_{w_1}, \ldots, z_{w_{d-2}} &\sim & t\left(\nu +d-1, 0, \frac{\left(\nu + y_{j_0}^2\right) \prod_{i=1}^{d-2}\left(\nu + i+ (z_{w_i})^2 \right)}{\prod_{i=1}^{d-1}(\nu +i)} \right) \nonumber \\
 & \stackrel{d}{=} &  t \left(\nu +d-1, 0, \frac{\nu + y_{j_0}^2+ \sum_{i=1}^{d-2} y_{w_i}^2}{\nu+d-1} \right) \, ,  \nonumber 
\end{eqnarray}  
because 
%----    
\begin{eqnarray} 
\frac{\left(\nu + y_{j_0}^2\right) \prod_{i=1}^{d-2}\left(\nu + i+ (z_{w_i})^2 \right)}{\prod_{i=1}^{d-1}(\nu +i)} &  \stackrel{(\ref{eq:eualrea22})}{=}  & \frac{\nu + y_{j_0}^2+ \sum_{i=1}^{d-3} y_{w_i}^2}{\nu+ d-2} \times  
\frac{\nu + d-2+ (z_{w_{d-2}})^2}{\nu +d-1} \nonumber \\
&=&    \frac{\nu + y_{j_0}^2+ \sum_{i=1}^{d-3} y_{w_i}^2}{\nu+ d-1} \times 
\left(1+ \frac{(z_{w_{d-2}})^2}{\nu +d-2}\right) \nonumber \\
& \stackrel{(\ref{eq:eualrea})}{=} &   \frac{\nu + y_{j_0}^2+ \sum_{i=1}^{d-3} y_{w_i}^2}{\nu+ d-1} \times \left(1+ \frac{y_{w_{d-2}}^2}{\nu + y_{j_0}^2+ \sum_{i=1}^{d-3} y_{w_i}^2}\right) \nonumber \\
&=&   \frac{\nu + y_{j_0}^2+ \sum_{i=1}^{d-2} y_{w_i}^2}{\nu+ d-1} \, . \nonumber
\end{eqnarray}   
The result holds using Remark \ref{rem:lcdepm}. 
   
%---------------------------------------- 
%----------------------------
\section{Proof of Proposition \ref{prop:cauco}} \label{app:theo:cauco}
%-------------- 
Knowing that for a multivariate Cauchy distribution $\bo{\X} \sim C_d(\boldsymbol{0}, \mathcal{I})$, $\X_{d} |\, x_{1}\, x_{2}\, \ldots, x_{d-1} \sim t\left(d, 0, \frac{1 + \sum_{k=1}^{d-1} x_{k}}{d} \right)$ (see \cite{devroye86}, Chapter 11), the recurrence reasoning leads to the following derivation.\\          
 First, for all $j_0 \in \{1,\ldots, d\}$, we have $\X_{w_1} |\, x_{j_0} = \sqrt{1 + x_{j_0}} \frac{Z_{w_1}}{\sqrt{2}} \sim t\left(2, 0, \frac{1 + x_{j_0}}{2}\right)$, and  \\ 
$\X_{w_2} |\, x_{j_0}\, z_{w_1} = \sqrt{h_1^2 + h_1 \frac{z_{w_1}}{\sqrt{2}}}\frac{Z_{w_2}}{\sqrt{3}}= \sqrt{1 + x_{j_0} + x_{w_1}} \frac{Z_{w_2}}{\sqrt{3}}  \sim t\left(3, 0, \frac{1 + x_{j_0} + x_{w_1}}{3}\right)$, which implies that $\X_{w_2} |\, x_{j_0}\, z_{w_1} \stackrel{d}{=} \X_{w_2} |\, x_{j_0}\, x_{w_1} \stackrel{d}{=} t\left(3, 0, \frac{1 + x_{j_0} + x_{w_1}}{3}\right)$. \\           
Second, we suppose that \\      
$
\X_{w_{d-2}} |\, x_{j_0}\, z_{w_1}\, \ldots, z_{w_{d-3}} = \sqrt{1 + x_{j_0} + \sum_{k=1}^{d-3} x_{w_k}} \frac{Z_{w_{d-2}}}{\sqrt{d-1}} \sim t\left(d-1, 0, \frac{1 + x_{j_0} + \sum_{k=1}^{d-3} x_{w_k}}{d-1} \right) 
$, with    
$  
h_{d-2} = \sqrt{1 + x_{j_0} + \sum_{k=1}^{d-3} x_{w_k}} $. \\   
Third,  we can write  
\begin{eqnarray}
\X_{w_{d-1}} |\, x_{j_0}\, z_{w_1}\, \ldots, z_{w_{d-2}} &=&
\sqrt{h_{d-2}^2 + h_{d-2} \frac{z_{w_{d-2}}}{\sqrt{d-1}}} \frac{Z_{w_{d-1}}}{\sqrt{d}} \nonumber \\ 
  &=& \sqrt{1 + x_{j_0} + \sum_{k=1}^{d-3} x_{w_k} + x_{w_{d-2}}} \frac{Z_{w_{d-1}}}{\sqrt{d}} \nonumber \\
  &=&  \sqrt{1 + x_{j_0} + \sum_{k=1}^{d-2} x_{w_k}} \frac{Z_{w_{d-1}}}{\sqrt{d}} \sim t\left(d, 0, \frac{1 + x_{j_0} + \sum_{k=1}^{d-2} x_{w_k}}{d} \right) \nonumber \, ,  
\end{eqnarray}     
 and the result holds.   

%%-------------------------------
%---------------------------
\section{Proof of Proposition \ref{prop:dirigp}} \label{app:prop:dirigp}
%--  
Let $D :=diag(R_j,\, R_{w_1}, \ldots, R_{w_{d-1}}) \in  \R^{d \times d}$ be a diagonal matrix. Since $(Y_1, \ldots, Y_d) \sim GD\left(\bo{a},\, \bo{b} \right)$ and bearing in mind the DM in (\ref{eq:dirig}), we can write
\begin{eqnarray} 
(Y_j, \bo{Y}_{\sim j}) &\stackrel{d}{=}&  \left(Y_{j},\, Z_{w_1} \left(1-Y_{j}\right), \ldots, Z_{w_{d-1}} \left(1-Y_{j}\right) \prod_{k=1}^{d-2}\left(1-Z_{w_k} \right) \right) \, , \nonumber \\
(Y_j^{1/p}, \bo{Y}_{\sim j}^{1/p}) &\stackrel{d}{=}&  \left(Y_{j}^{1/p},\, \left[Z_{w_1} \left(1-Y_{j}\right)\right]^{1/p}, \ldots, \left[Z_{w_{d-1}} \left(1-Y_{j}\right) \prod_{k=1}^{d-2}\left(1-Z_{w_k} \right)\right]^{1/p} \right) \, .   \nonumber
\end{eqnarray} 
 Thus, we have
\begin{eqnarray}
& & (\X_{j},\,\bo{\X}_{\sim j}) \stackrel{d}{=} D \left[Y_j^{1/p}, (\bo{Y}_{\sim j}^{1/p})^\T \right]^\T \nonumber \\  
& & \stackrel{d}{=}  \left(R_j Y_{j}^{1/p},\, R_{w_{1}}\left[Z_{w_1} \left(1-Y_{j}\right)\right]^{1/p}, \ldots, R_{w_{d-1}}\left[Z_{w_{d-1}} \left(1-Y_{j}\right) \prod_{k=1}^{d-2}\left(1-Z_{w_k} \right)\right]^{1/p} \right) \, , \nonumber \\   
& & \stackrel{d}{=}  \left(\X_{j},\, R_{w_{1}}\left[Z_{w_1} \left(1- |\X_{j}|^p \right)\right]^{1/p}, \ldots, R_{w_{d-1}}\left[Z_{w_{d-1}} \left(1-|\X_{j}|^p\right) \prod_{k=1}^{d-2}\left(1-Z_{w_k} \right)\right]^{1/p} \right) \, , \nonumber  
\end{eqnarray}   
because $\X_{j} \stackrel{d}{=} R_j Y_{j}^{1/p}$ implies that $ |\X_{j}| \stackrel{d}{=}  Y_{j}^{1/p}$ and $ |\X_{j}|^p \stackrel{d}{=}  Y_{j}$.  

%---------------------------------------------------
%-----------------------------------
\section{Proof of Corollary \ref{coro:idconstga}} \label{app:theo:idconstga}   
%-----  
For Point (i), if $Y_i := \X_i^c/c$ then $Y_i \sim \Gamma\left(a_i,\, \beta/c \right)$ and the constraint becomes $\sum_{i=1}^d Y_i =1$. Therefore, we know that a subset of $d-1$ variables of $(Y_{j}, \, Y_{w_1}, \ldots, Y_{w_{d-1}})$ follows the  Dirichlet distribution, that is, $(Y_{j}, \, Y_{w_1}, \ldots, Y_{w_{d-2}}) \sim D(a_{j},\, a_{w_1}, \ldots, a_{w_{d-1}})$ (see \cite{devroye86}, Chapter 11, Theorem 4.1). The last variable is given by $Y_{w_{d-1}} = 1 - Y_{j} - \sum_{i=1}^{d-2} Y_{w_{i}}$.        
Thus, we use Equation (\ref{eq:diri}) to model $(Y_{j}, \, Y_{w_1}, \ldots, Y_{w_{d-2}})$ and the inverse transformation $X_i^c = c \, Y_i$ to obtain the result bearing in mind that $W \sim Beta(a, \, b) \Longrightarrow  c\, W \sim B1(c,\, a,\, b)$. \\ 
For Point (ii), if we use the following transformation $Y_i = G_i^{-1} \circ F_{\X_i}\left(X_i \right)$, then $Y_i \sim \Gamma(a_i,\beta)$ with $i=1, \ldots, d$ and  $Y_i,\, i=1, \ldots, d$  are independent. Moreover, the constraint becomes
 $\sum_{i=1}^{d} Y_i =c$. Using Point (i) and the inverse transformation of the form $F_{\X_i}^{-1}\circ G_i$, we obtain the result.  
 
%------------------------------------ 
%---------------------------
\section{Proof of Corollary \ref{coro:idconstga2}}   \label{app:theo:idconstga2}
%----  
For Point (i), consider a random variable $\X_{d+1} \sim \Gamma(a_{d+1} =1, \,\beta)$, which is independent of $(\X_i,\, i=1,\,\ldots,\, d)$ such that $\sum_{i=1}^{d+1} \X_i = c$. If we use 
$Y_i = \frac{X_i}{\sum_{k=1}^{d+1} \X_k} = \frac{X_i}{c}$, then it is known that  
$\left(Y_{j}, Y_{w_1}, \ldots, Y_{w_{d-1}} \right) \sim D(a_{j},\, a_{w_1},\, \ldots, a_{w_{d-1},\, 1})$, which implies that $Y_{j} + Y_{w_1} + \ldots + Y_{w_{d-1}} < 1$ or equivalently 
$\X_{j} + \X_{w_1} + \ldots + \X_{w_{d-1}} < c$ (see \cite{devroye86}, Chapter 11, Theorem 4.1). Thus, we use Equation (\ref{eq:diri}) to model $(Y_{j}, \, Y_{w_1}, \ldots, Y_{w_{d-1}})$ and the inverse transformation $X_i^c = c \, Y_i$ to obtain the result. \\ 
For Point (ii), if we use the following transformation $Y_i = G_i^{-1} \circ F_{\X_i}\left(X_i \right)$, then $Y_i \sim \Gamma(a_i,\beta)$; $Y_i,\, i=1,\, \ldots, d$  are independent, and the constraint becomes $\sum_{i=1}^{d} Y_i < c$. Using Point (i) and the inverse transformation $F_{\X_i}^{\leftarrow}\circ G_i$, we obtain the result.
         
%------------------------------------       
%------------------------
\section{Proof of Theorem \ref{theo:idconstg}} \label{aap:theo:idconstg}
%----
If $\X_i \sim \mathcal{N}(0, \, \sigma_i^2),\, i=1,\ldots, d$ such that $\sum_{i=1}^{d} \X_i = c$, it is known that a subset of $d-1$ variables of that random vector follows the multivariate normal distribution with mean $\boldsymbol{\mu} := \left(\frac{c \sigma_i^2}{\sum_{k=1}^d\sigma_k^2},  \, i=1,\,\ldots,\, d-1 \right)$ and covariance $\Sigma^c$ (see \cite{vrins18} for more details). \\    
For Point (i),  we choose $\bo{\X}^c_{d-1} =(\X_{j}^c,\, \X_{w_i}^c, \, i=1,\ldots, d-2)$ as that subset of variables, and  we have  $\bo{\X}^c_{d-1} \sim \mathcal{N}(\boldsymbol{\mu}, \, \Sigma^c)$. The last variable is given by $\X_{w_{d-1}}^c = c- \X_{j}^c- \sum_{i=1}^{d-2} \X_{w_i}^c$. The Point (i) holds by applying Equation (\ref{eqn:gdis}). \\ 
For Point (ii), if we use the transformation $Y_i = \sigma_i \Phi^{-1} \circ F_{\X_i}\left(\X_i \right)$, then $Y_i \sim \mathcal{N}(0,\, \sigma_i^2)$; $Y_i,\, i=1,\, \ldots, d$  are independent and the constraint becomes $\sum_{i=1}^{d} Y_i = c$. Using Point (i) and the inverse transformation, we obtain the result.      
 
%-------------------------------------- 
%--------------------------------------
\section{ Proof of Corollary \ref{coro:idconstgau2}} \label{app:coro:idconstgau2}
%----  
If we use $Y_i:=\X_i^2$, then $Y_i \sim \Gamma(\alpha_i=1/2, \beta=1/2)$ with $i=1,\, \ldots,\, d$, and the constraint becomes $\sum_{i=1}^d Y_i= c$. We then use Corollary \ref{coro:idconstga} to obtain a DM of $Y_i, \, i=1,\, \ldots,\, d$ under the constraint $\sum_{i=1}^d Y_i= c$. Point (i) holds because $X_j \stackrel{d}{=} R_j |\X_j| = R_j \sqrt{Y_j}$ with $Y_j \sim B1(c, 1/2, (d-1)/2)$ and the fact that $W \sim Beta(a,\, b)$ implies $\sqrt{W} \sim GB1(2,\, 1,\, a,\, b)$ and $ r\, \sqrt{W} \sim GB1(2,\, r,\, a,\, b)$. \\          
Point (ii) holds by using the transformation $Y_i = \Phi^{-1} \circ F_{\X_i}\left(\X_i \right)$ and Point (i). 
     
%-------------------------------
\end{appendices}   			
  
%-----------------------------------------------------------------
%----------------------------------------------------------------
\section*{References}         
% \bibliographystyle{elsarticle-num}         
% \bibliography{bibi_tsi}  

%%%%%%%%%%%%%%%%%%%%%%%%%%%%%%%%%%%%%%%%%%%%%%%%%%%%%%%%%%%%%%
%%%%-----------------------------------------------------
\end{document}